\documentclass[preprint,3p,12pt]{elsarticle}

\usepackage{graphicx, appendix}
\usepackage[colorlinks=true]{hyperref}
\usepackage{amssymb, bm, amsmath}
\usepackage{wrapfig}
\usepackage{float}
\usepackage{multirow}
\usepackage{array}
\usepackage{booktabs}
\usepackage{xcolor}
\usepackage{lipsum}
\usepackage{placeins}
\usepackage{float}
\biboptions{numbers,sort&compress}

\journal{a journal for review}

\begin{document}

\begin{frontmatter}


\title{Coupled Bend-Twist Mechanics of Biomimetic Scale Substrate}

\author[1]{Sanjay Dharmavaram}
\ead{sd045@bucknell.edu}
\author[2]{Hossein Ebrahimi}
\ead{ebrahimi@knights.ucf.edu}
\author[2]{Ranajay Ghosh \corref{cor1}\corref{cor2}}
\ead{ranajay.ghosh@ucf.edu}

\cortext[cor1]{Corresponding author address: 4000 Central Florida Blvd, Orlando, Florida 32816, USA}
\cortext[cor2]{Corresponding author telephone: +1 407-823-3402}
\address[1]{Department of Mathematics, Bucknell University, Lewisburg, PA}
\address[2]{Department of Mechanical and Aerospace Engineering, University of Central Florida, Orlando, FL}

\begin{abstract}

We develop the mechanics of combined bending and twisting deformation of a one-dimensional filamentous structure with protruding stiff fish-scale-like plates embedded at an angle on the surface. We develop Cosserat kinematic formulation along with scale contact constraints. This geometrically exact model allows us to bypass the limitations of typical finite element computations inherent in these systems when deflections are large. The derived structure-property relationships reveal for the first time the combined effect of bending and twisting on a slender fish scale inspired substrate. The model subsumes previous models on pure bending and twisting but also shows previously unobserved phenomena that arise due to the coupled effects of these loads. This includes a new interpretation of kinematic locking behavior, multiple contact regimes, asymmetric sensitivities of one curvature over the other, and sharp transitions in the nonlinear moment-curvature and torque-twist behaviors reflecting the complex scale engagement patterns.

\end{abstract}




\end{frontmatter}


\section{Introduction}
\label{Intro}
\begin{figure}[ht]
 \centering
 \begin{tabular}{cc}
 \includegraphics[width=2.8in]{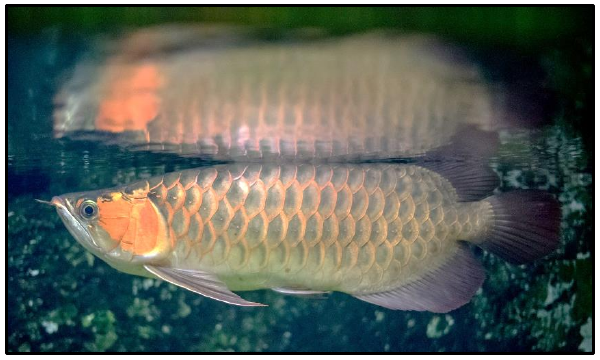} & 
 \includegraphics[width=2.8in]{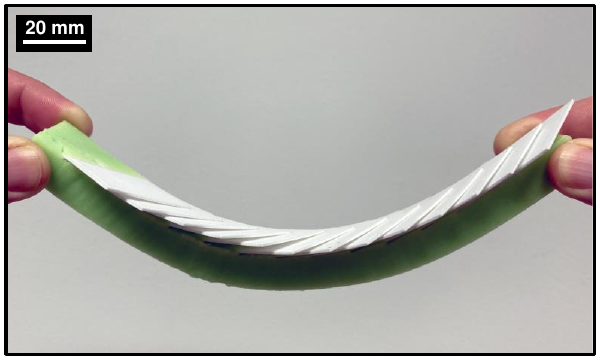}\\
 (a) & (b)\\
 \includegraphics[width=2.8in]{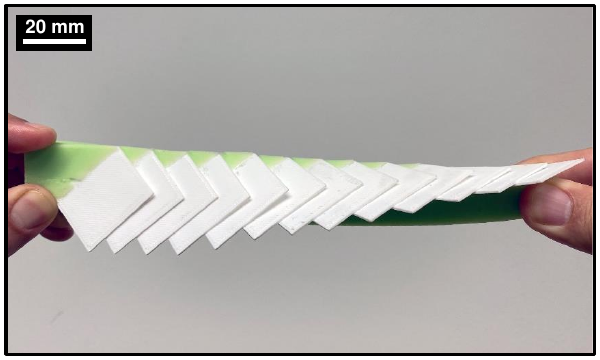} &
 \includegraphics[width=2.8in]{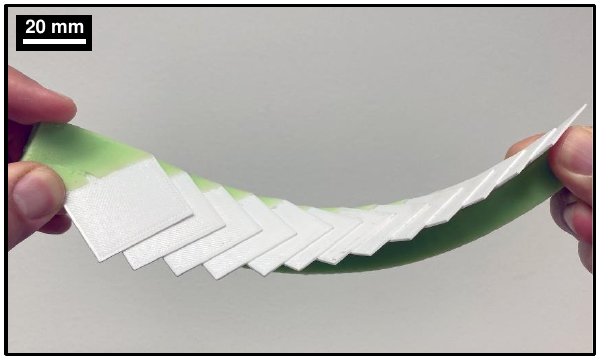}\\
 (c) & (d)
 \end{tabular}
 \caption{(a) Natural fish scales under deformation mode, adapted under CC BY 2.0 \cite{kemoole2017}. (b) Fabricated biomimetic scale metamaterials under bending deformation. (c) Fabricated biomimetic scale metamaterials under twisting deformation. (d) Fabricated biomimetic scale metamaterials under combined bending and twisting deformation.}
 \label{fig:fish}
\end{figure}

Fishes are synonymous with scales, Fig.~\ref{fig:fish}(a), although scales are far more versatile in nature. They  cover numerous reptiles and can also be intermittently found in mammals such as in pangolins and armadillos \cite{wang2016pangolin,chen2011armadillo}. More interestingly, there are scale-like features in the wings of butterflies, human hair and papillae on feline tongues \cite{latorre2006investigation,michielsen2008gyroid} indicating the singular importance of the scale morphology in enhancing functions. One of the major advantages of scales architecture is that they are are generally lightweight additions to a substrate due to low volume fraction and yet enhance stiffness, and multifunctionality \cite{buehler2006nature,wegst2015bioinspired}. Several critical properties including protection, locomotion, camouflaging, and thermal regulation have been attributed to scales \cite{kertesz2008photonic,long1996functions,song2010quantitative}. Thus, they are now intensely studied as material templates to make armors, smart skins, soft robotics and multifunctional surfaces \cite{sadati2015stiffness,wei2016novel,roche2017soft,sire2009origin}.  Mechanically, scales give rise to fascinating emergent behavior such as strain stiffening, evolving directionality and anomalous frictional response \cite{ghosh2014contact,ghosh2016frictional,ghosh2017non,ali2019bending,ali2019tailorable,ali2019frictional,ali2020tailorable,ebrahimi2019tailorable,ebrahimi2020coulomb,ebrahimi2021emergent,ebrahimi2021fish}. These behaviors can potentially aid organisms in balancing multiple complex and often contradictory functions such as locomotion with protection, and softness with stiffness. 

Such possibilities have resulted in numerous studies in the past to understand the nature of property enhancements brought about by scales. Early research highlighted and confirmed the outstanding behaviors of 1-dimensional beam-like substrates that were covered uniformly with scales using a combination of analytical and finite element (FE) models \cite{ghosh2014contact,ebrahimi2019tailorable,vernerey2010mechanics}. These works  established precise structure-property relationships in pure bending loads for both smooth and rough sliding between scales, and for both rigid and flexible scales \cite{ghosh2014contact,ghosh2016frictional,vernerey2010mechanics,vernerey2014skin}. The essential characteristics of bending behavior such as strain stiffening and locked states were found to be universally valid even when stiff scales were not uniformly distributed (e.g. functionally graded \cite{ali2019tailorable}) or loaded under non-uniform bending \cite{ali2019bending}. The locked state is essentially a kinematic configuration beyond with scale motion would result in interpenetration leading to rigid behavior. In reality, somewhat before locking commences, the contact forces on the scales would be large enough to lead to scales deformation. Thus, bending behavior would transition from substrate deflection (soft) to scale deformation(stiff), leading to a sharp increase in rigidity of the overall structure.  Recent studies have further confirmed locking and nonlinear strain stiffening phenomena under pure torsion, thereby extending their validity even in twisting \cite{ebrahimi2019tailorable,ebrahimi2020coulomb}. 
However, torsion structure-property calculations while underlining the universality also highlighted the striking differences from bending. For instance, the kinematic locking envelopes turned out to be a complex nonlinear functions of geometry, unlike the bending case where they were linear. In addition, the tilt angle of the scales, Fig. \ref{fig:fish} (b-d) had a significant impact on the nature of locking, with some angles even precluding  locking behavior. When Coulomb friction was included between the sliding scales, more differences with the bending case were evident \cite{ghosh2016frictional,ebrahimi2020coulomb}. For instance, frictional locking was indeed common between bending and twisting, but in twisting, the frictional locking envelopes were highly nonlinear without closed form solution unlike the bending case. Also, the relative energy dissipated from friction  during one cycle of loading (start to lock) was found to monotonically increase with the friction coefficient ($\mu$) for twisting case, even though for bending case, increasing $\mu$ did not always increase  dissipated work in a cycle \cite{ebrahimi2020coulomb,ghosh2016frictional}. Such differences are well anticipated as the mechanical behaviors are dependent on the interplay of structure, load and geometry with no intrinsic guarantees of universality. Thus, it is important to investigate individual canonical load cases carefully and rigorously before declaring either generalities or anomalies. 

Surprisingly, the mechanics of combined bending and twisting has not been studied. There is virtually no knowledge of the behavior of slender fish scale inspired substrates in 3-D or spatial deflections, which is significantly more complex but of great practical utility due to the nature of real world loads and possibility of geometrical defects in the biomimetic substrate, leading to cross curvatures (e.g. sagging or intrinsic twist). Prior research outlined above show that existing models may not be universally valid, scale linearly or even hold similar functional forms across geometry-loading combinations. Thus entirely new models are necessary to arrive at structure-property relationships. 

We address this lacuna by developing the mechanics of a frictionless biomimetic scale-covered beam of the geometric form shown in Fig. \ref{fig:fish} (b-d). We use a Cosserat rod model to reveal the nature and regimes of nonlinearity, the interplay of bending and twisting, relationships between global (substrate) and local (scale) deformations, and locking behavior. In the process, we obtain the structure-property relationships. We use FE simulations and prior results in the literature to validate our model. This paper is organized as follows: in Sec. \ref{sec:engagement}, we develop the kinematics of the filament in bending and twisting loads, and derive the moment-curvature relationships, assuming small strains and additive strain energies. In Sec. \ref{sec:fem}, we briefly describe the finite element model we use to validate our theory. In Sec. \ref{sec:results}, we conclude the paper with results and discussion.

\section{Contact Mechanics of Scales Engagement}
\label{sec:engagement}
We develop a kinematic description of the scale-covered substrate whose schematic is shown Fig. \ref{fig:schematic} (a). The challenge in describing kinematics of this system is the complicated three-dimensional configurations of the scales and their rotation with respect to the substrate. A Cosserat kinematic description, commonly used to model the finite elasticity of rods, is ideal in the present context. Although the structure shown in Fig. \ref{fig:schematic} has  finite width, any variation of global strains along the width (i.e., along the x-direction) has been neglected. 
We therefore model the substrate as an idealized rod. Since Cosserat kinematics incorporates a natural director basis at every point along the rod, the orientation of the scale relative to the substrate is easy to describe using this basis. In this way, the scales' orientations naturally couple to the bending strain variables of the Cosserat description. Another advantage of the Cosserat framework is that modeling the mechanics of these structures becomes straightforward. We add the strain energies of the substrate and the homogenized energy for scales; see Sec. \ref{sec:mechanics}. This work will assume the substrate's constitutive law to be linearly elastic which assumes small material strains in the substrates. This has been found to be sufficient in the practical contexts of relatively thin substrates that tend to lock before very large strains are permitted. More complicated constitutive models can be easily incorporated \cite{antman1995nonlinear, audoly2010elasticity} within our framework. Other simplifying assumptions and the general validity of the energy formulation have been discussed elsewhere\cite{ebrahimi2019tailorable,ghosh2017non}. 

\subsection{Global Cosserat Kinematics}
\label{Global Kinematics}
\begin{figure}[htbp]
 \centering
 \begin{tabular}{cc}
 \includegraphics[width=3.0in]{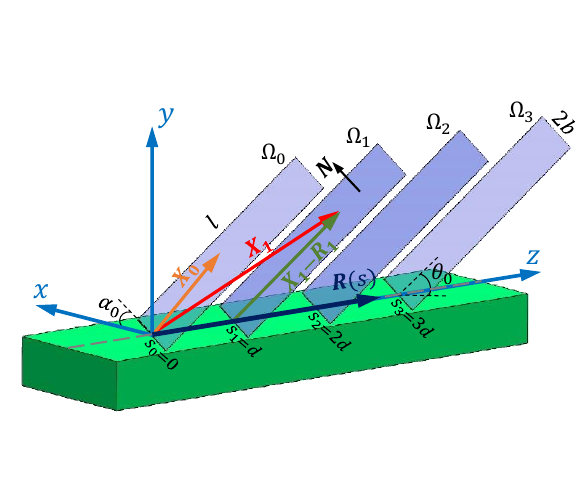} & 
 \includegraphics[width=3.0in]{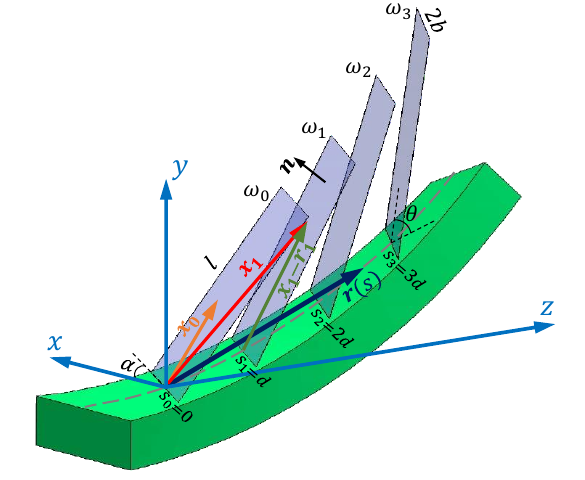}\\
 (a) & (b)
 \end{tabular}
 \caption{(a) Schematic of flat reference configuration of the scale-covered rod with arc length variable along the length. (b) Schematic of deformed configuration of the same rod under coupled bend-twist load.}
 \label{fig:schematic}
\end{figure}

\noindent
Consider the schematic of the system shown in Fig. \ref{fig:schematic} (a). We model the substrate as a Cosserat rod \cite{antman1995nonlinear,audoly2010elasticity} whose  centroidal curve in its undeformed reference configuration is defined by $\mathbf{R}(s)$, where $s$ represents the arc-length along the undeformed configuration. As shown in the figure, the planar undeformed substrate points along the z-direction, i.e., $\mathbf{R}(s) = s\mathbf{e}_3$, where $\mathbf{e}_1$, $\mathbf{e}_2$, and $\mathbf{e}_3$ are the standard Cartesian basis vectors along the $x$, $y$, and $z$ directions, respectively. We identify the reference directors of the undeformed rod ($\mathbf{D}_i,\;i=1,2,3$) with the cartesian basis vectors, i.e., $\mathbf{D}_i=\mathbf{e}_i$. Let $\mathbf{r}(s)$ denote the deformed position of the centroidal curve (shown in Fig. \ref{fig:schematic} (b)), and $\mathbf{d}_1(s)$, $\mathbf{d}_2(s)$, and $\mathbf{d}_3$, the orthonormal directors moving along the deformed rod. In the Cosserat description, the directors of the deformed configuration are related to those of the undeformed configuration via an orthogonal (rotation) matrix $\mathbf{Q}(s)$:
\begin{equation}
\mathbf{d}_{i}(s) = \mathbf{Q}(s)\mathbf{e}_i, \text{ for }i=1,2,3.
\label{eq:d=Qe} 
\end{equation} 

Differentiating \eqref{eq:d=Qe} with respect to $s$ and substituting for $\mathbf{e}_i$ using the same, we obtain
\begin{equation}
\mathbf{d}_i'(s) = \mathbf{K}\mathbf{d}_i,\text{ for }i=1,2,3,
\label{eq:d_prime_Q}
\end{equation} 
where 
\begin{equation}
  \mathbf{K}:=\mathbf{Q}'(s)\mathbf{Q}^T(s)
  \label{eq:K=Q'QT}
\end{equation}
is a skew-symmetric matrix associated with the bending and twisting strains in the rod.  We can associate $\mathbf{K}$ (with respect to the $\{\mathbf{d}_1,\mathbf{d}_2,\mathbf{d}_3\}$ basis) with an axial vector, $\bm{\kappa}:=\kappa_1\mathbf{d}_1+\kappa_2\mathbf{d}_2+\kappa_3\mathbf{d}_3$, i.e.,
\begin{equation}
\mathbf{K} = \left(
\begin{array}{ccc}
0 & -\kappa_3 & \kappa_2\\ \kappa_3 & 0 & -\kappa_1\\ -\kappa_2 & \kappa_1 & 0 \end{array} \right), \label{eq:Kkappa} 
\end{equation} 
and write \eqref{eq:d_prime_Q} as
\begin{equation}
    \mathbf{d}_i' = \bm{\kappa}\times\mathbf{d}_i\text{ for }i=1,2,3.
\end{equation}
The strain variable $\kappa_3$ is interpreted as the twisting strain in the rod, while $\kappa_1$ and $\kappa_2$ as the bending strains for bending about the $x$ and $y$ axes, respectively. Note that the definition of bending and twisting strains in the Cosserat sense described above is distinct from the actual 3D strains in the substrate. In Cosserat kinematics, the rod is modeled as a 1D curve, and Cosserat strains can be interpreted as the  (3D) strains averaged over the cross-section of the substrate~\cite{antman1995nonlinear}. This approximation is justified for thin structures. In the cases of pure bending \cite{ghosh2014contact} or pure twisting \cite{ebrahimi2019tailorable}, we identify the Cosserat strains with the twist rates and curvature of the rod, respectively. However, the matrix description of the strains noted in \eqref{eq:Kkappa} is appropriate in the general case considered here. \emph{Henceforth in this paper, we use the term `strains' strictly in this Cosserat sense}.  

We assume spatially homogeneous strain in this model analogous to prior studies~\cite{ghosh2014contact, vernerey2010mechanics}. This makes $\mathbf{K}$ independent of $s$ and helps set up periodicity conditions to extract structure-property relationships. For sharp gradients in strains or functionally graded structures, either local periodicity~\cite{ali2019tailorable} or discrete scale by scale approach can be used~\cite{ali2019bending}. The periodicity condition allows us to study the kinematics using a representative volume element (RVE) of a pair of scales~\cite{ghosh2014contact,vernerey2010mechanics}.



The strain variables $\kappa_1$, $\kappa_2$, and $\kappa_3$ are viewed as parameters. Equation \eqref{eq:K=Q'QT} can be explicitly integrated to solve for $\mathbf{Q}(s)$ and we obtain
\begin{equation}
\mathbf{Q}(s) = e^{s\mathbf{K}},
\label{eq:Qsol} 
\end{equation} 
where without loss of generality we have assumed that the scale at $s=0$ is fixed and does not change its orientation as the scales deform. That is, $\mathbf{Q}(0)=\mathbf{I}$. 

Besides bending and twisting, the rod would more generally experience shearing in $x$ and $y$ directions, and stretching in the $z$-direction (we neglect other stretching and shear modes due to the substrate's thin cross-section). In  Cosserat kinematics, strain variables associated with the above three modes of deformation are given by $\nu_1, \nu_2$, and $\nu_3$, respectively. These are related to the deformation by
\begin{equation}
\mathbf{r}'(s) = \sum_{i=1}^3\nu_i(s)\mathbf{d}_i(s).
\label{eq:r'=nu d}
\end{equation} 
If we use \eqref{eq:d=Qe} to express $\mathbf{d}_i$ in terms of $\mathbf{Q}(s)$ and $\mathbf{e}_i$, and after integrating the resulting equation, we obtain
\begin{equation}
 \mathbf{r}(s) = \sum_{i=1}^3\int_0^s \nu_i\mathbf{Q}(\tau)\mathbf{e}_i\;d\tau,
 \label{eq:r=int}
\end{equation}
where we have used the boundary condition $\mathbf{r}(0)=\mathbf{0}$, that is, the rod is fixed at the origin $s=0$. In this work, we assume inextensibility and unshearability of the rod and set $\nu_1=\nu_2=0$ and $\nu_3=1$. These are reasonable approximations for the current case as soft polymers tend to be incompressible and the cross sections are not known to distort significantly. In the FE simulations presented in Sec. \ref{sec:fem}, we approximate inextensibility by choosing the Poisson's ratio $\nu=0.49$.  Numerical simulations presented in Sec. \ref{sec:results} seem to confirm this approximation, as we observe no appreciable length change of the rod under the combined loading. Neither is any significant change in cross-sectional area observed. Note that in spite of these assumptions, there is little loss of generality. If the model presented were to be applied for design and optimization of compressible substrates, the above assumptions could easily be relaxed. In that case, the position of the rod must be determined using \eqref{eq:r=int} with appropriate values for parameters $\nu_1$, $\nu_2$, and $\nu_3$.

\subsection{Local Kinematics of Scales}
\noindent
We assume that the scales are identical in shape (rectangular with exposed length $l$ and width $2b$), equally spaced along the rod, and are much stiffer than the underlying substrates and thus assumed to be rigid. In addition, the assumption of periodicity of contact (i.e., the use of RVEs) is also common~\cite{ghosh2014contact,ebrahimi2019tailorable}. Non-periodic or functionally graded systems have been investigated previously for the cases of pure bending and pure twisting and have indicated the salient features of these systems are still preserved under periodicity~\cite{ali2019bending, ali2019tailorable}.  In their reference configuration, the scales are oriented parallel to each other with the orientation described in terms of the dihedral angles $\alpha_0$ and $\theta_0$; see Fig.~\ref{fig:schematic} (a). Here, $\theta_0$ is the angle made by the scale with the $z$-axis, and $\alpha_0$ is the scale's tilt angle made by its base with the $x$-axis.  The separation of two adjacent scales measured along the centroidal curve of the undeformed configuration is $d$. As the rod deforms, the scales rigidly rotate with the rod (maintaining orientations $\theta_0$ and $\alpha_0$ with the deformed centerline) until contact between adjacent scales occurs. When the scales are under contact, they continue to remain rigid, but the contact induces a change in their orientations, \emph{viz.}, to $\theta$ and $\alpha$ due to scale rotation on the substrate. 

We now derive conditions for contact between adjacent scales. Let $s_i$ represent the position of the scale $i$ along the arc-length of the undeformed rod. Let $\Omega_i\subset\mathbb{R}^3$ be the set of all points constituting scale $i$ in the reference configuration, and $\omega_i\subset\mathbb{R}^3$ be the set of all points on the same, but in the deformed configuration. These are shown schematically in Figs.~\ref{fig:schematic} (a) and (b), respectively. The midpoint of the base of the scales are located at $\mathbf{R}_i := \mathbf{R}(s_i)$ in the reference configuration and at $\mathbf{r}_i := \mathbf{r}(s_i)$ in the deformed configuration. Note that for prescribed strains $\kappa_i$ ($i=1,2,3$), $\mathbf{r}(s_i)$ is given by \eqref{eq:r=int} where $s_i=i\cdot d$. It is clear from Fig. \ref{fig:schematic} that  for any given $\mathbf{X}_i\in\Omega_i$ and $\mathbf{x}_i\in\omega_i$, the vectors $\mathbf{X}_i-\mathbf{R}_i$ and $\mathbf{x}_i-\mathbf{r}_i$ lie on $\Omega_i$ and $\omega_i$, respectively. Since $\mathbf{N}$ is normal to the scale $i$ (in its reference configuration), it follows that
\begin{equation}
\mathbf{N}\cdot[\mathbf{X}_i-\mathbf{R}_i]=0. 
\end{equation} 
Since the scales rotate rigidly under deformation, we have
\begin{equation}
 \mathbf{x}_i-\mathbf{r}_i = \mathbf{Q}_i\left[\mathbf{X}_i-\mathbf{R}_i\right],
 \label{eq:scale-deformation-relation}
\end{equation}
where $\mathbf{Q}_i:=\mathbf{Q}(s_i)=\mathbf{Q}(i\cdot d)$ is the rotation matrix given by \eqref{eq:Qsol} evaluated at $s=i\cdot d$. Two adjacent scales (say, $i=0$ and $i=1$) in their deformed configurations intersect if and only if they have at least one common point. That is, if and only if the following equation has at least one solution:
\begin{equation}
\mathbf{x}_1-\mathbf{x}_0=\mathbf{0}.
\label{eq:intersection condition}
\end{equation} 
Using \eqref{eq:scale-deformation-relation}, the previous equation can be equivalently written as
\begin{subequations}
\begin{equation}
 \mathbf{r}_1 + \mathbf{Q}_1[\mathbf{X}_1-\mathbf{R}_1] -\mathbf{X}_0=\mathbf{0},
\end{equation}
where we have used $\mathbf{Q}_0=\mathbf{I}$, and $\mathbf{R}_0=\mathbf{r}_0=\mathbf{0}$ to account for the boundary condition of scale $i=0$. Since $\mathbf{X}_0$ and $\mathbf{X}_1$ lie on (finite) planes $\Omega_0$ and $\Omega_1$, respectively, these points must satisfy
\begin{equation}
\mathbf{N}\cdot\mathbf{X}_0=0,\text{ for }\mathbf{X}_0\in\Omega_0,
\label{eq:plane1}
\end{equation}
\begin{equation}
\mathbf{N}\cdot\Big[\mathbf{X}_1-\mathbf{R}_1\Big]=0, \text{ for }\mathbf{X}_1\in\Omega_1. 
\label{eq:plane2}
\end{equation} 
\label{eq:intersection condition summary}
\end{subequations}

\begin{figure}[htbp]
 \centering
 \includegraphics[width=3.0in]{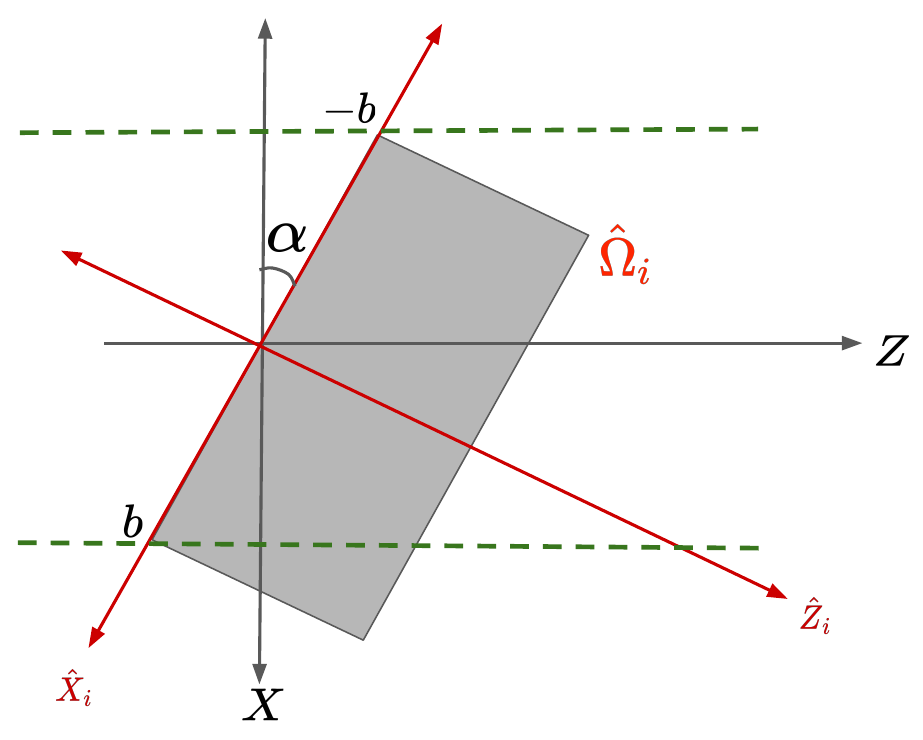}
 \caption{Schematic showing the coordinate system for $\hat{\mathbf{X}}_i$. The dashed lines represent the edges of the substrate and the gray region is the projection of the scale on the $X-Z$ plane. For each scale $i$, matrix $\mathbf{T}$ is a coordinate transformation from $XYZ$ to $\hat{X}_i\hat{Y}_i\hat{Z}_i$.}
 \label{fig:scale_rotation}
\end{figure}

Note that solutions, $\mathbf{X}_0$ and $\mathbf{X}_1$, of \eqref{eq:intersection condition summary} are the 3D coordinates of the points of intersection of the scales, pulled back to the reference configuration. From a computational viewpoint, while it is easy to solve this linear system of equations, verifying if the solutions lie on \emph{finite} planes $\Omega_0$ and $\Omega_1$ is not as straightforward. This is because of two factors---finiteness of $\Omega_1$ and $\Omega_2$, and their complicated 3D orientation in $\mathbb{R}^3$. We deal with both of them by employing the following change of coordinates:
\begin{equation}
\hat{\mathbf{X}}_i := \mathbf{T}[\mathbf{X}_i-\mathbf{R}_i],\;i=0,1,
\label{eq:new_vars}
\end{equation}
where $\hat{\mathbf{X}}_0$ and $\hat{\mathbf{X}}_1$ are rotated coordinate variables and 
\begin{equation}
    \mathbf{T} = \left(
    \begin{array}{ccc}
    \cos\alpha & 0 & \sin\alpha\\
    0 & 1 & 0\\
    -\sin\alpha & 0 & \cos\alpha
    \end{array}
    \right),
\end{equation}
is a rotation matrix. The transformation rule \eqref{eq:new_vars} maps $\Omega_i$ to $\hat{\Omega}_i:=\mathbf{T}(\Omega_i)$, where the latter's projection on the $\hat{X}_i$-$\hat{Z}_i$ plane is shown in gray in Fig. \eqref{fig:scale_rotation}. Thus,  $\mathbf{T}$ rotates the scales such that  
\begin{equation}
 \hat{\Omega}_0=\hat{\Omega}_1=\mathcal{B} := [-b,b]\times[0,\infty]\times[0,l\cos\theta].
 \label{eq:boundB}
\end{equation}

Rewriting \eqref{eq:intersection condition summary} in terms of $\hat{\mathbf{X}}_0$ and $\hat{\mathbf{X}}_1$ and using \eqref{eq:new_vars}, we obtain the following equivalent conditions for the intersection of the two adjacent scales:
\begin{subequations}
\begin{equation}
\mathbf{T}^T\hat{\mathbf{X}}_0 - \mathbf{Q}_1\mathbf{T}^T\hat{\mathbf{X}}_1 = \mathbf{r}_1,
\label{eq:system1a}
\end{equation}
\begin{equation}
 \mathbf{N}^T\mathbf{T}^T\hat{\mathbf{X}}_0=0,
\label{eq:system1b}
\end{equation}
\begin{equation}
 \mathbf{N}^T\mathbf{T}^T\hat{\mathbf{X}}_1=0,
\label{eq:system1c}
\end{equation}
\label{eqs:system1}
\end{subequations}
where $\hat{\mathbf{X}}_0, \hat{\mathbf{X}}_1\in\mathcal{B}$ and superscript $T$ denotes matrix transpose. Thus, transformed variables $\hat{\mathbf{X}}_i$ lie in a simple rectangular domain $\mathcal{B}$. Transforming the variables in this manner also resolves the issue of finiteness noted above because it is straightforward to check if solutions lie in $\mathcal{B}$ by checking the bounds on the variables.

\subsection{Contact Conditions}
\begin{figure}[htbp]
 \centering
 \includegraphics[width=6.0in]{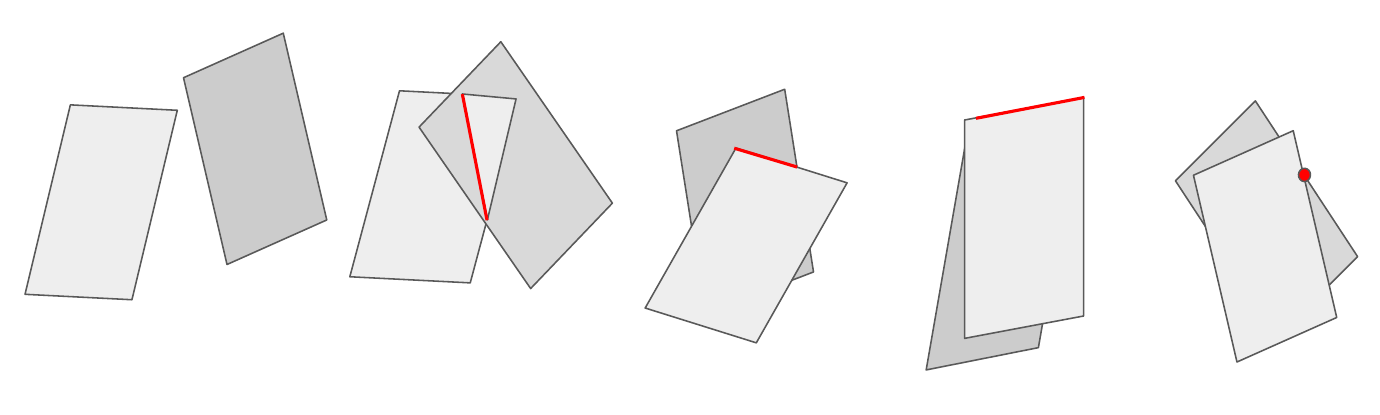}
 \caption{Various possibilities for two finite rectangles to intersect. The left-most figure shows no contact between scales, the next shows interpenetration of scales. The last three are a non-exhaustive sampling of possible modes of contact. Intersections are shown in red.}
 \label{fig:modes of intersection}
\end{figure}

Observe that \eqref{eqs:system1} is an under-determined linear system with five equations in six unknowns. If $\hat{\mathbf{X}}_0$ and $\hat{\mathbf{X}}_1$ remain unrestricted (i.e., $\hat{\mathbf{X}}_i\in\mathbb{R}^3$, instead of $\mathcal{B}$), then we have two possibilities---no solutions, when the two planes containing the scales are parallel, or infinitely many solutions with a line of intersection. However, since the scales constitute finite planes, since $\hat{\mathbf{X}}_0$ and $\hat{\mathbf{X}}_1$ are restricted to $\mathcal{B}$. When finite planes intersect, there are the following possibilities: a) no solution, when the planes are either parallel or do not intersect within the domain, b) infinitely many solutions when the two planes interpenetrate each other with a line-segment intersection c) \emph{unique or infinitely} many solutions (depending on the relative orientation of the scales), but the scales \emph{do not interpenetrate} each other. The third case is of interest here as they correspond to physically realizable configurations of scales making contact. A schematic for these configurations is shown in Fig. \ref{fig:modes of intersection} with intersections (if any) shown in red.

There are various ways in which the third case of intersecting yet non-penetrating scales may be realized. Observe that in any of these configurations there is \emph{at least one point of intersection} that falls in the following two cases: 1) it is either a corner of one of the scales (\emph{corner contact}) or 2) is a point of intersection of two edges of different scales (\emph{edge contact}). See last three figures in Fig. \ref{fig:modes of intersection}). To find the intersection points for the first  case, we prescribe corner coordinates $\hat{X}_i$ and $\hat{Z}_i$ (for scale $i$) and for the second case, any two coordinates but from different scales, e.g., $\hat{X}_i$ and $\hat{Z}_j$ ($i\neq j$) are prescribed. We use the bounds for the corresponding variables appearing in \eqref{eq:boundB} to fix these values. Since $\hat{Y}_i$ is unbounded we do not fix this coordinate. The two cases are not necessarily mutually exclusive, as some configurations can be both. For instance, when corner contact happens on an edge; see third schematic in Fig. \ref{fig:modes of intersection}. The cases described above do not automatically exclude interpenetrating configurations. For example, the second schematic in Fig. \ref{fig:modes of intersection} is an edge-contact. In computations, we exclude such cases by explicitly checking for interpenetration. 

There are eight possible corner contacts depending on which one of the eight edges (between the two scales) makes contact with the other. And there are sixteen edge contacts depending which one of the four edges of first scale makes contact with the four of the other. Of these twenty four possibilities most correspond to configurations with extreme twists that lead to appreciable material nonlinearities or even cause the scales to debond from the substrate. Therefore, we only consider the following four cases, shown schematically in Fig.~\ref{fig:schematic different modes of contact}:

\begin{figure}[htbp]
    \centering
    \begin{tabular}{cccc}
         \includegraphics[width=1.55in]{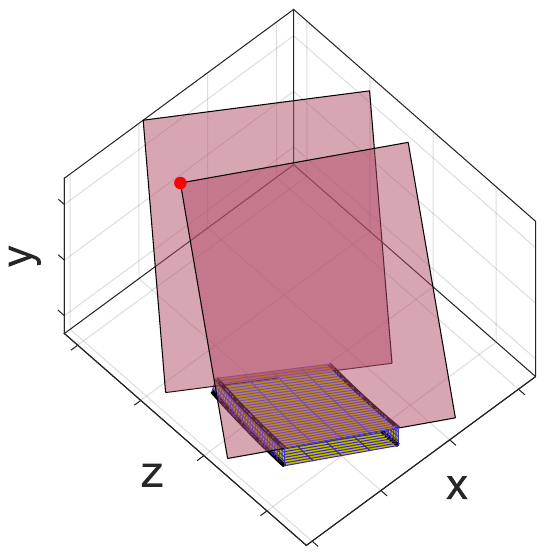} & 
         \includegraphics[width=1.55in]{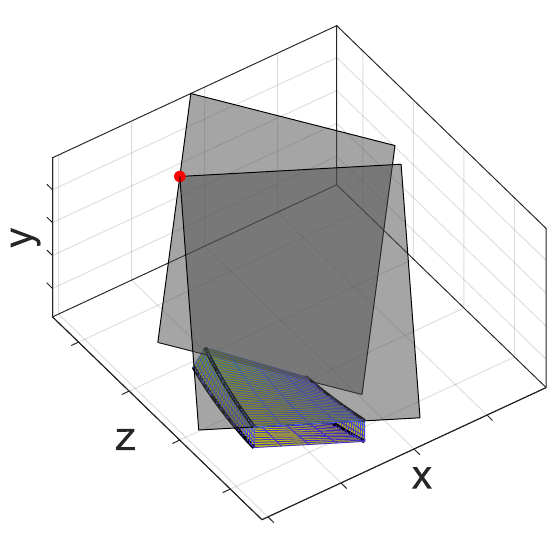}    & 
         \includegraphics[width=1.55in]{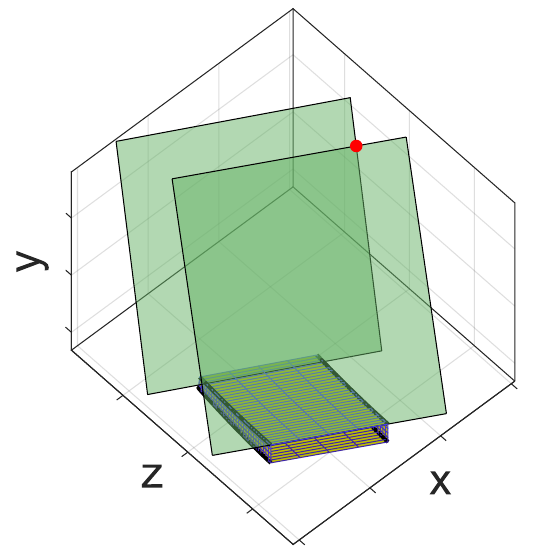} &
         \includegraphics[width=1.45in]{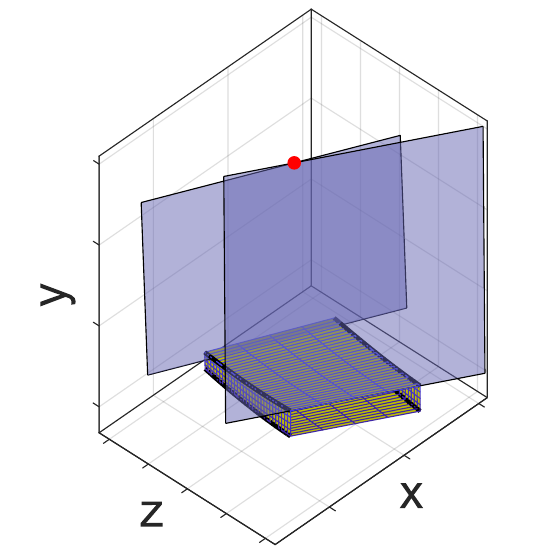}
         
         \\
         (a) & (b) & (c) & (d) 
    \end{tabular}
    \caption{Modes of contact for an RVE considered in this work: (a) Corner contact (red) (b) Top-left edge contact (black) (c) Top-right edge (green) (d) Top edge (blue). Contact point is highlighted as a red dot}
    \label{fig:schematic different modes of contact}
\end{figure}
\begin{enumerate}
    \item Corner contact: Either of the two top corners of scale `$0$' makes contact with scale `$1$'. For such a configuration, we set $(\hat{X}_0,\hat{Z}_0)=(\pm b,l\cos\theta)$. This is shown as red in our schematic Fig.~\ref{fig:schematic different modes of contact} (a). In practice, $\hat{X}_0=-b$ is found only when $\alpha<0$.
    \item Top-Left edge contact: This happens when the top edge of scale `0' makes contact with left edge of scale `1'. For this, we set $(\hat{X}_1,\hat{Z}_0)=(- b,l\cos\theta)$. This is shown as black in our schematic; see Fig.~\ref{fig:schematic different modes of contact} (b).
    \item Top-Right edge contact: This happens when the top edge of scale `0' makes contact with left edge of scale `1'. For this, we set $(\hat{X}_1,\hat{Z}_0)=( b,l\cos\theta)$. This is shown as green in our schematic; see Fig.~\ref{fig:schematic different modes of contact} (c).
    \item Top-edge contact: This happens when the top edge of scale `0' makes contact with top edge of scale `1'. For this, we set $(\hat{Z}_0,\hat{Z}_1)=( l\cos\theta,l\cos\theta)$. This is shown as blue in our schematic; see Fig.~\ref{fig:schematic different modes of contact} (c).
\end{enumerate}
These color codes will be used later in Sec. \ref{sec:results} . All the above cases can be expressed mathematically as the conditions
\begin{subequations}
\begin{equation}
\mathbf{a}_0^T \hat{\mathbf{X}}_0 + \mathbf{a}_1^T\hat{\mathbf{X}}_1 = A,
\label{eq:a_0}
\end{equation}
\begin{equation}
\mathbf{b}_0^T\hat{\mathbf{X}}_0 + \mathbf{b}_1^T\hat{\mathbf{X}}_1= B,
\label{eq:a_1}
\end{equation}
\label{eqs:a_0,a_1}
\end{subequations}
where $\mathbf{a}_0,\;\mathbf{a}_1,\;\mathbf{b}_0,\;\mathbf{b}_1$ are 3D vectors that select the appropriate coordinates of $\hat{\mathbf{X}}_0$ and $\hat{\mathbf{X}}_1$ that we wish to fix, and $A$ and $B$ are the bounds on the coordinates.  For example, for the corner contact case, i.e., Case 1, noted above with   $(\hat{X}_0,\hat{Z}_0)=(b,l\cos\theta)$, \eqref{eqs:a_0,a_1}, we have $\mathbf{a}_0=(1,0,0)\;\mathbf{a}_1=(0,0,0)^T$ and $\mathbf{b}_0=(0,0,1),\;\mathbf{b}_1=(0,0,0)^T$ (which pick coordinates $\hat{X}_0$ and $\hat{Z}_0$, respectively), and $A=b$, $B=l\cos\theta$. For Case 2, $(\hat{X}_0,\hat{X}_1)=(b,b)$, so $\mathbf{a}_0=(1,0,0),\;\mathbf{a}_1=(0,0,0)^T$, $\mathbf{b}_0=(0,0,0),\;\mathbf{b}_1=(1,0,0)^T$, $A=b$, and $B=b$.

To obtain an implicit expression for the dependence of $\mathbf{K}$ (and therefore, $\kappa_1$, $\kappa_2$ and $\kappa_3$) on $\theta$, we solve \eqref{eqs:system1} simultaneously with \eqref{eqs:a_0,a_1}. We do this in the following steps. First, we solve \eqref{eq:system1a} for $\hat{\mathbf{X}}_1$, i.e., $\hat{\mathbf{X}}_1=\mathbf{T}\mathbf{Q}_1^T\mathbf{T}^T\hat{\mathbf{X}_1}-\mathbf{T}\mathbf{Q}_1^T\mathbf{r}_1$. We then plug the previous expression for $\hat{\mathbf{X}}_1$ into \eqref{eq:a_0} and \eqref{eq:a_0} to obtain equations only involving $\hat{\mathbf{X}}_0$. Taking the resulting equations together with \eqref{eq:system1b}, we obtain a $3\times 3$ system of linear equations $\mathbf{A}\hat{\mathbf{X}}_0=\mathbf{c}$, where
\begin{equation}
 \mathbf{A} = \left(
 \begin{array}{c}
 (\mathbf{T}\mathbf{N})^T\\
 (\mathbf{a}_0+\mathbf{T}\mathbf{Q}_1\mathbf{T}^T\mathbf{a}_1)^T\\
 (\mathbf{b}_0+\mathbf{T}\mathbf{Q}_1\mathbf{T}^T\mathbf{b}_1)^T
 \end{array}
 \right),\;\mathbf{c} = \left(
 \begin{array}{c}
 0\\
 A + \mathbf{a}_1^T\mathbf{T}\mathbf{Q}_1^T\mathbf{r}_1\\
 B + \mathbf{b}_1^T\mathbf{T}\mathbf{Q}_1^T\mathbf{r}_1\\
 \end{array}
 \right).
\end{equation}

Plugging the solution $\hat{\mathbf{X}}_0=\mathbf{A}^{-1}\mathbf{c}$ along with the solution $\hat{\mathbf{X}}_1=\mathbf{T}\mathbf{Q}_1^T\mathbf{T}^T\hat{\mathbf{X}_1}-\mathbf{T}\mathbf{Q}_1^T\mathbf{r}_1$ in \eqref{eq:system1a} we obtain the following implicit dependence of $\theta$ in terms of $\mathbf{K}$:
\begin{equation}
 f(\mathbf{K},\theta) = \mathbf{N}^T\mathbf{Q}_1^T\mathbf{T}^T\mathbf{A}^{-1}\mathbf{c}-\mathbf{N}^T\mathbf{Q}_1^T\mathbf{r}_1=0.
 \label{eq:f(K,theta)=0}
\end{equation}

The dependence of  $f$ on $\mathbf{K}$ and $\theta$ can be gleaned by observing that $\mathbf{Q}_1$ and $\mathbf{r}_1$ depend on $\mathbf{K}$ ((cf. \eqref{eq:Qsol} and \eqref{eq:r=int}), while $\mathbf{N}$, $A$ and $B$ depend on $\theta$. Parameters $\alpha$ appears in  $\mathbf{T}$, and $l$ and $b$ appear in $A$, and $B$. Equation \eqref{eq:f(K,theta)=0} is a highly nonlinear `bridging' law linking the local to the global kinematics completing the multiscale kinematics description of the system. We solve this equation numerically to obtain the dependence of $\theta$ on the bending strains, $\kappa_i$. Solutions corresponding to interpenetrating configurations are discarded.


\subsection{Mechanics of Biomimetic Scale Slender Substrate}
\label{sec:mechanics}
In order to understand the mechanics of this structure, we take recourse to energy balance between the global loads and local deformation. We model the underlying substrate as an Euler elastica whose strain energy is given by:

\begin{equation}
   \mathcal{E}_{beam} = \int_0^L \Big[\frac{1}{2}B_1 \kappa_1^2 + \frac{1}{2} B_2\kappa_2^2+ \frac{1}{2}D\kappa_3^2\Big]\;ds.
\end{equation}
where $B_1,B_2$ and $D$ are respectively the bending and twisting rigidities. For beams with circular cross sections, $B_1=EI_1, B_2=EI_2$ and $D=GJ$ where $E$ is the Young's modulus, $I_1,I_2$ are respective area moments, $G$ is the shear modulus, and $J$ is the polar moment. In our case, two complications arise. The embedding of the scales lead to an inclusion effect on the beam, resulting in additional stiffness (composite effect). For bending deformation, we postulate a scaling parameter $C_f$ to the bending rigidities, i.e. $B_1=C_fEI_1, B_2=C_fEI_2$. The values of the parameter $C_f$ can be obtained using FE simulations~\cite{ebrahimi2019tailorable}. For the twisting deformation, a non-circular cross section also introduces warping effect. Although warping is not substantial to affect the kinematics of the scales engagement, it does have an effect on the energy. This is addressed using another multiplicative factor $C_w$, which is readily available in handbooks~\cite{ugural2011advanced} and often scaled with respect to $EI_1$ rather than $GJ$. Thus, the twisting rigidity takes the form $D=C_fC_wEI_1$. 

To include the contribution from the scales, we note that a scale's rotations is captured by the angles $\theta$ and $\alpha$. As the substrate deforms under applied strain, the scales rotate freely until a critical threshold of curvatures is reached when the scales engage. Let $\Gamma_e\subset \mathbb{R}^3$ denote the bending strains $(\kappa_1,\kappa_2,\kappa_3)$, for which the scales are engaged. This region can be determined by solving \eqref{eq:f(K,theta)=0} for $\kappa_1$, $\kappa_2$ and $\kappa_3$ for which $\theta>\theta_0$ and when scales make contact. We determined it numerically. Clearly, the region will depend on the parameters chosen. In Fig.~\ref{fig:engagement region} of Sec. \ref{sec:results}, we show a plot for a slice of $\Gamma_e$ along $\kappa_2=0$. Since the scales are built into the substrate, under engagement (i.e., when scales make contact), their rotations are resisted by the substrate. We model the substrate resistance using linear torsional springs with the elastic energy stored in the springs for each scale given by:
\begin{equation}
    \mathcal{E}_{scale}(\theta,\alpha) = \frac{1}{2d}\Big[K_\theta(\theta-\theta_0)^2+K_\alpha(\alpha-\alpha_0)^2\Big]H_{\Gamma_e}(\kappa_1,\kappa_2,\kappa_3),
    \label{eq:escale}
\end{equation}
where $K_\theta$ and $K_\alpha$ are the spring constants of the torsional springs, and $H_{\Gamma_e}$ is the indicator function on $\Gamma_e$ (i.e., $H_{\Gamma_e}(\kappa_1,\kappa_2,\kappa_3)=1,\text{ if }(\kappa_1,\kappa_2,\kappa_2)\in\Gamma_e$, and zero, otherwise). The total energy per RVE can be additively written as: 
\begin{equation}
 \mathcal {E}(\kappa_1,\kappa_2,\kappa_3)=\mathcal{E}_{beam}(\kappa_1,\kappa_2,\kappa_3)+ \mathcal{E}_{scale}(\theta,\alpha).
 \label{eq:energy per RVE}
\end{equation}
Our FE simulations show that the change in $\alpha$ from $\alpha_0$ is minimal even when the scales are engaged. This observation is in agreement with our earlier findings for the cases of pure bending and twisting \cite{ghosh2014contact, ebrahimi2019tailorable}. We therefore drop its dependence from $\mathcal{E}$ and henceforth fix $\alpha$ to $\alpha_0$. Note that in light of \eqref{eq:f(K,theta)=0}, $\theta$ is itself a function of $(\kappa_1,\kappa_2,\kappa_3)$, hence the dependence of $\mathcal{E}$ on the same.

$K_\theta$ is related to the Young's modulus of the substrate ($E$), scale thickness ($t_s$), inclusion length $(L)$ and $\theta_0$. As we have shown in \cite{ghosh2014contact,ebrahimi2019tailorable} the following non-dimensional scaling exists:
\begin{equation}
    \frac{K_\theta}{E t_s^2}=C_B \left(\frac{L}{t_s}\right)^nf(\theta_0),
\end{equation}
where $n$ is non-dimensionless constant that we estimate using $FE$ simulations, $C_B$ is a constant, and $f(\theta_0)$ is function of angle $\theta$. For the results presented below, these were $C_B=3.62$, $n=1.55$, $f(\theta_0)\approx1$ \cite{ebrahimi2019tailorable,ebrahimi2020coulomb}. 

The moments in the three directions are computed by differentiating \eqref{eq:energy per RVE} with respect to $\kappa_1$, $\kappa_2$, and $\kappa_3$:
\begin{subequations}
\begin{equation}
M_1 = C_f{E}I_1\kappa_1 + \frac{K_\theta}{d}(\theta-\theta_0)\frac{\partial\theta}{\partial\kappa_1}H_{\Gamma_e}(\kappa_1,\kappa_2,\kappa_3),
\end{equation}
\begin{equation}
M_2 = C_f{E}I_2\kappa_2 + \frac{K_\theta}{d}(\theta-\theta_0)\frac{\partial\theta}{\partial\kappa_2}H_{\Gamma_e}(\kappa_1,\kappa_2,\kappa_3),
\end{equation}
\begin{equation}
 M_3 = C_f C_w GI_1 \kappa_3+ \frac{K_\theta}{d}(\theta-\theta_0)\frac{\partial\theta}{\partial\kappa_3}H_{\Gamma_e}(\kappa_1,\kappa_2,\kappa_3).
\end{equation}
\label{eqs:moments}
\end{subequations}

These relations can be computed numerically to obtain the moment-curvature relationships. Note that the derivative of the indicator function, $H_{\Gamma_e}$ is the (surface) Dirac Delta distribution which is zero everywhere except on the boundary of $\Gamma_e$. This term does not appear in \eqref{eqs:moments} because it is multiplied by $\theta-\theta_0$, which regularizes to zero precisely on the boundary of $\Gamma_e$.

\section{Finite Element Analysis} 
\label{sec:fem}

We build Finite Element (FE) models on ABAQUS/CAE 2017 (Dassault Syst\`emes) to validate the analytical model developed above. In the simulations, we model  the substrate and the scales as 3D deformable solids. The substrate is a rectangular prismatic member 
(length $L_B=200$ mm) onto which a row of 19 identical scales are embedded on one side. In this assembly, scales are spaced $d=10$ mm apart, oriented with angle of $\theta_0=5^\circ$ with respect to the substrate's top surface, and angle of $\alpha_0=30^\circ$ with respect to the substrate’s rectangular cross section. Since our analytical model applies to an RVE, where strains are constant, we chose our substrate to be sufficiently long so as to avoid edge effects. The substrate was modeled as a linear elastic material with an elastic modulus $E_B=2.5$ MPa and Poisson's ratio $\nu=0.49$. Thus, the shear modulus of the substrate is $G_B={\frac{E_B}{2(1+\nu)}}= 0.84$ MPa. Scales were modeled to be rigid with respect to the substrate by imposing rigid body constraints. The contact between the rigid scales is modeled using the 'Surface to Surface Contact' algorithm of the software~\cite{abaqus20146}. This contact option is typically used in literature for biomimetic scale problems of this type~\cite{ghosh2014contact,ali2019bending,ali2019tailorable,ebrahimi2019tailorable}.

The mechanical loads for bending and twisting were applied quasi-statically to the system and as boundary conditions. In our numerical studies, we set $\kappa_2=0$. That is, we do not explore bending in the transverse direction to the substrate, which is considered trivial in the current context. To compare the combined effect of bending (with strain $\kappa_1$) and twisting (with strain $\kappa_3$) with the semi-analytical model presented above, we perform the simulation in two static steps with nonlinear geometry option (NLGEOM on) \cite{abaqus20146}. In the first step, the bending rotations with the same magnitude but opposite directions were applied to the cross-sections at either ends. The magnitude of bending was increased linearly from 0 to (approximately) $\kappa_1 L_B /2$, where $L_B$ is the length of the substrate.  The final value could be attained only approximately as edge effects precluded a constant bending strain could not be maintained throughout the substrate. To ameliorate these effects we extracted an averaged strain in the beam far from the ends. More details on this procedure will be explained in \ref{AppendixA}. In the second step, the bending rotations was fixed at the final value ($\kappa_1L_B/2$) and the twisting rotations were applied to the both end cross-sections with reverse directions, again linearly increasing from 0 to (approximately) $\kappa_3 L_B /2$ during the step time. A frictionless surface-to-surface contact was applied to the scales surfaces. To test the reliability of the numerical results, we carried out a mesh convergence study. Sufficient mesh density was found for different regions of the model confirming computational accuracy. Mesh convergence led to a total number of about 230,000 elements. Because of the complex geometry of the system particularly due scales inclusions in the substrate, the top layer of substrate was meshed with tetrahedral quadratic elements C3D10 and other regions were meshed with quadratic hexahedral elements C3D20. 

\section{Results and Discussion}
\label{sec:results}

In this section we present results exploring cross-curvature effects, with various types of scales contact color-coded differently in the plots. Without loss of generality, we take the initial scale angle $\theta_0=0^\circ$ (grazing scales) since we are more interested in scale engagement behavior. A $\theta_0>0$ would only shift the plots forward till engagement occurs, with the curves thereafter following the same trajectories~\cite{ghosh2014contact, ebrahimi2019tailorable}. We use the following non-dimensional parameters: $\eta=l/d$, $\beta=b/d$, and $\hat{\kappa}_i=\kappa_i d$, $i=1,2,3$ to explore the effects.

\begin{figure}[ht]
 \centering
   \includegraphics[width=6.5in]{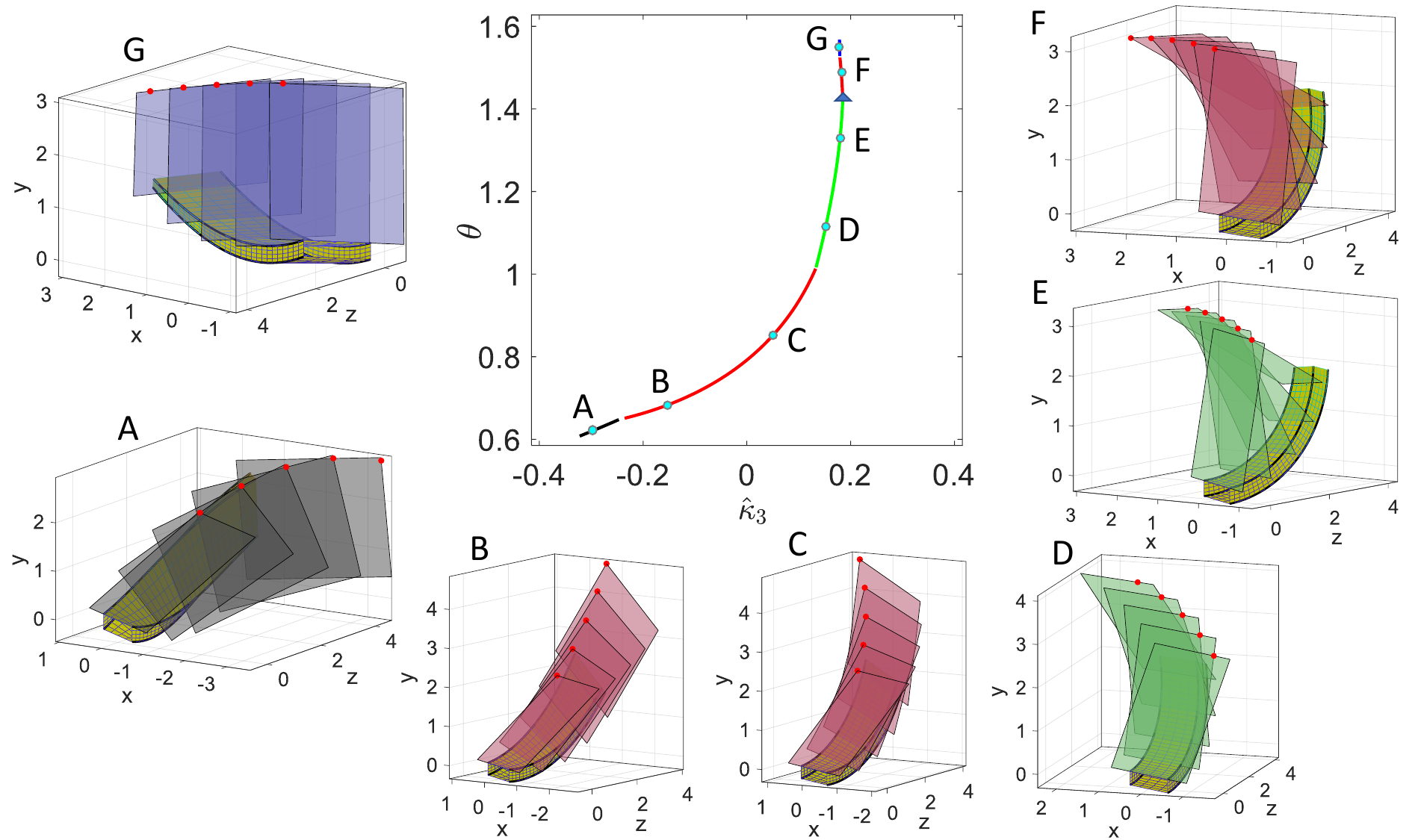}
 \vspace{-10pt}
 \caption{$\hat{\kappa}_3$ vs $\theta$ with different modes of contact. Here the fixed parameters are: $\hat{\kappa}_1 = 0.23$, $\eta=3$, $\beta=1.25$ and $\alpha=30^\circ$.}
 \label{fig:different modes of contact}
\end{figure}

A typical $\hat{\kappa}_3$  (twist) versus $\theta$ (scale angle) behavior (for a given $\hat{\kappa}_1$) is shown in Fig.~\ref{fig:different modes of contact}. In this particular figure, we have used $\hat{\kappa}_1=0.23$, $\eta=3$, $\beta=1.25$ and $\alpha=30^\circ$. The presence of an initial bending can lead to additional scale contact types, not observed for pure twisting investigated previously~\cite{ebrahimi2019tailorable,ebrahimi2020coulomb}. These various contact regimes are color coded in the twist-angle plot as follows:  the black curve is the contact type where the top edge of the first scale makes contact with the left edge of the second,  the red curve represent configurations where the corner of the first scale makes contact with the face of the second scale, green curve the configuration where the top edge of the first scale makes contact with the right edge of the second scale, and the blue curve represents the configuration where the top edges of the two scales make contact. We also provide animations of the contact kinematics (see supplementary material) to aid in better visualization of this system. 

Recall that in the case of pure twisting \cite{ebrahimi2019tailorable} (or pure bending \cite{ghosh2014contact}) the kinematic locking is typically identified in the strain(curvature/twist)-angle plot at the point where the slope goes to infinity. However, in the case of combined load, at locking, the slopes changes sign abruptly from positive to negative. This critical point is indicated by a blue triangle in Fig.~\ref{fig:different modes of contact} where the curve is predicted to transition from green to red. However, in reality configurations beyond this critical point are physically impossible without violating interpenetration condition. This critical point is thus interpreted as locking in this case. This non-orthogonal condition of locking is a unique artifact of combined load system, not found in pure bending or twisting~\cite{ghosh2014contact,ebrahimi2019tailorable}. In the subsequent analysis, we treat post-locking configurations as forbidden. For the parameter values explored in this work, we find that the blue curves (which correspond to configurations with top edge contact) always occur as locked states. Hence, they are not explicitly shown in the figures below. In addition, contact regimes that correspond to very high twisting strains are also not considered as they are of little practical significance and difficult to achieve without introducing significant material nonlinearity. 

\begin{figure}[h!]
    \centering
\begin{tabular}{cc}
 \includegraphics[width=2.9in]{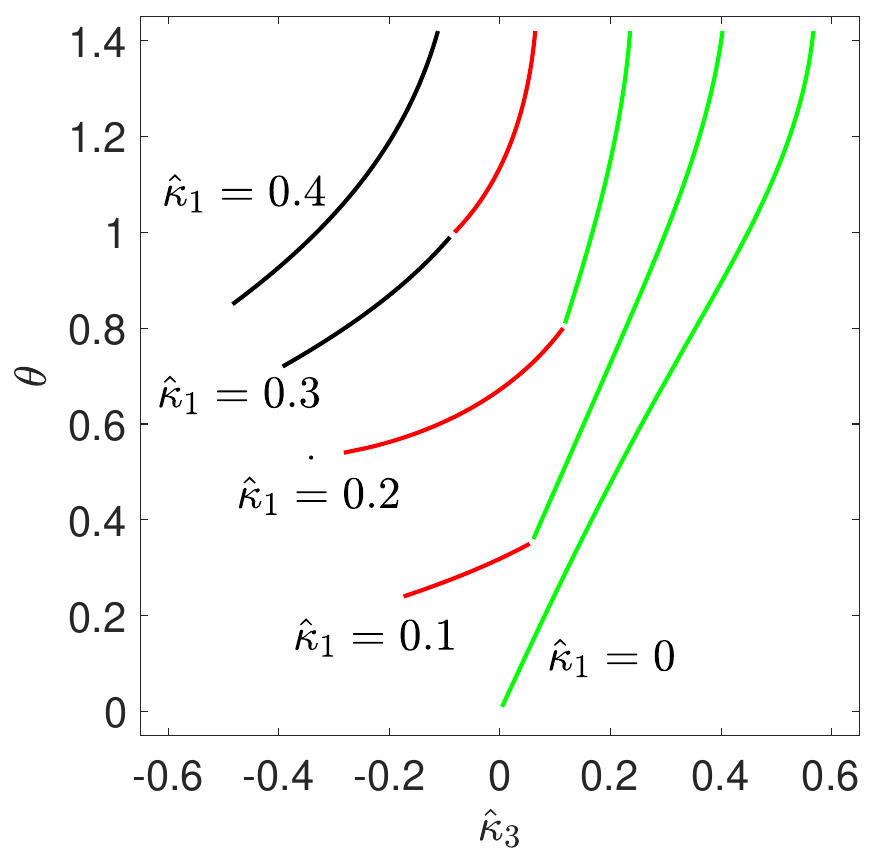} 
 &  \includegraphics[width=2.9in]{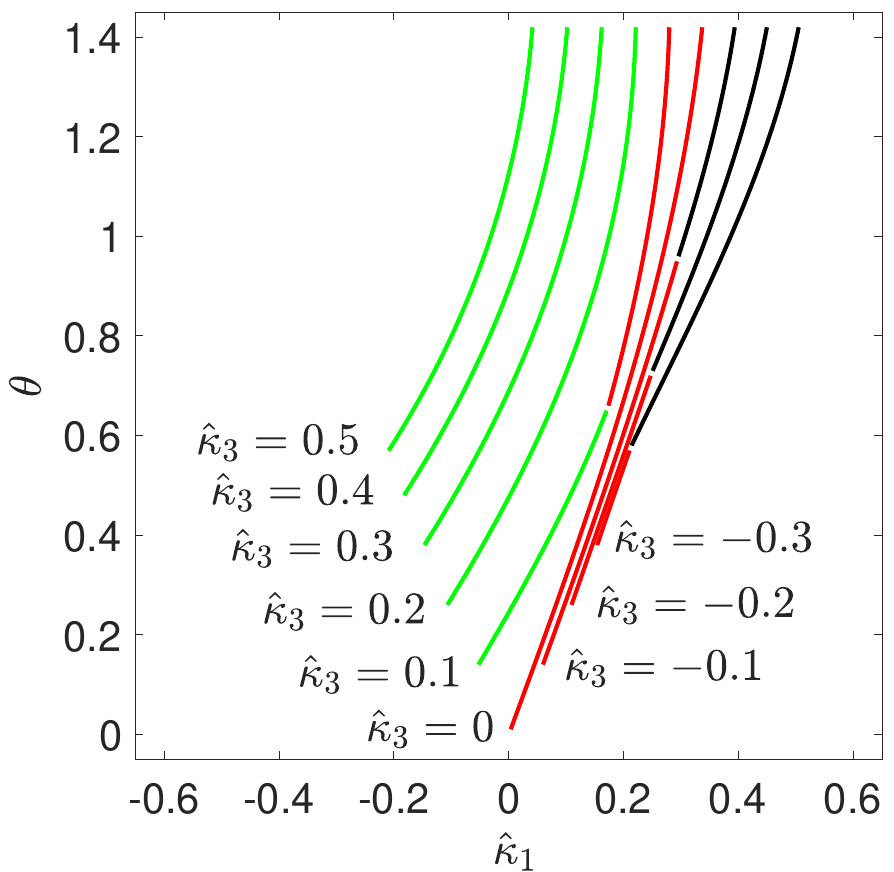} \\
  (a)   & (b) \\
 \includegraphics[width=2.9in]{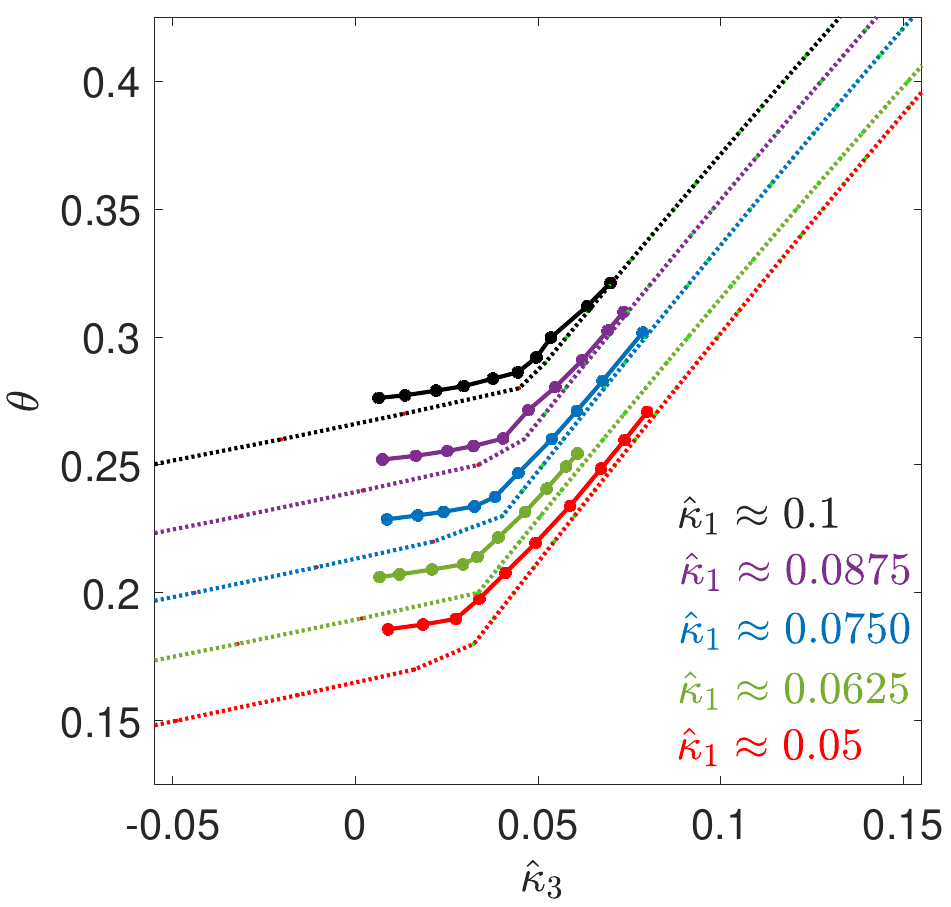} &
 \includegraphics[width=2.9in]{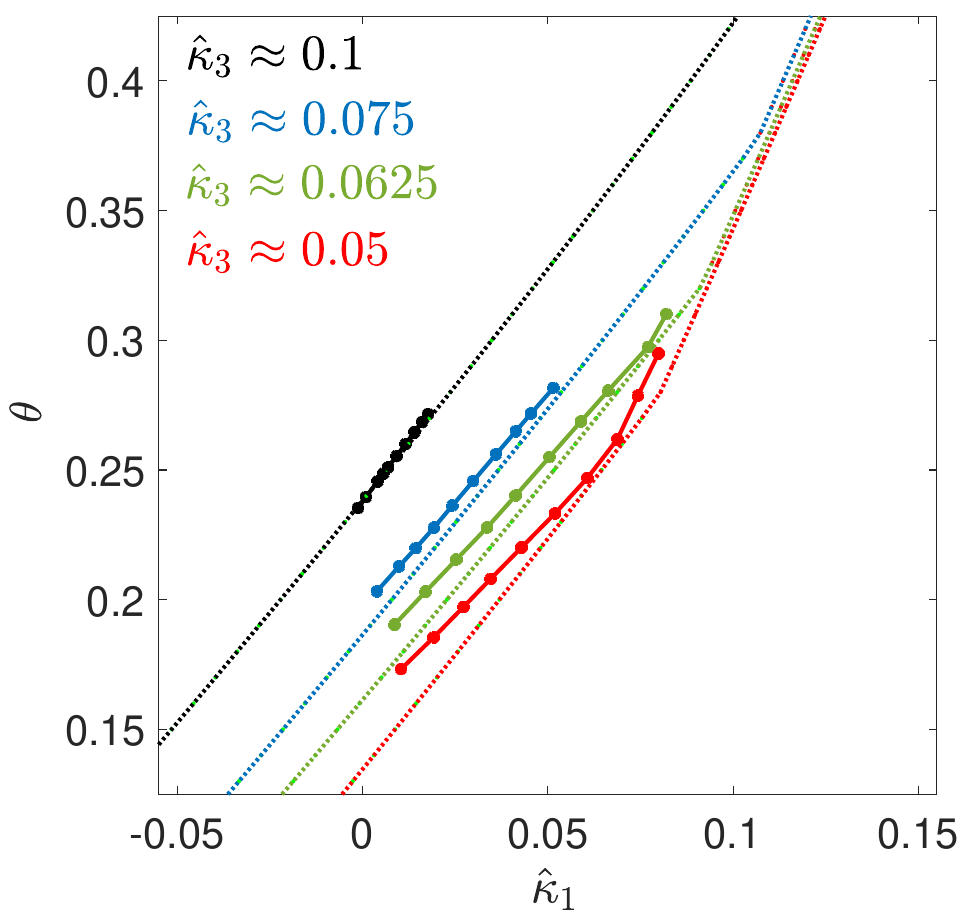}\\
 (c) & (d)
\end{tabular}
\caption{(a) $\theta$ versus $\hat{\kappa}_3$ for different values of bending strains ($\hat{\kappa}_1$). (b) $\theta$ versus $\hat{\kappa}_1$ for different values of twisting strains ($\hat{\kappa}_3$). (c) FE simulations of  $\theta$ versus $\hat{\kappa}_3$ for different values of bending strains ($\hat{\kappa}_1$) (d) FE simulations of  $\theta$ versus $\hat{\kappa}_1$ for different values of twisting strains ($\hat{\kappa}_1$), $\eta=3$, $\beta=1.25$ and $\alpha=30^\circ$.} 
\label{fig:kappa31 vs theta different kappa13}
\end{figure}

We now explore in detail the cross coupling effects of one curvature over the other. In Fig.~\ref{fig:kappa31 vs theta different kappa13} (a) we show the effect of bending ($\hat{\kappa}_1$) on the twisting ($\hat{\kappa}_3$) versus scale angle ($\theta$). The various values of $\hat{\kappa}_1$ are shown beside their corresponding curves. This figure is fundamentally different from pure bending~\cite{ghosh2014contact} or twisting~\cite{ebrahimi2019tailorable} kinematics. Here, there are distinct regions of kinematics and they are dependent on the existing bending curvature, with distinct contact regimes emerging throughout engagement. Thus existence of bending can result in entirely new type of kinematic behaviors in twisting. The color coded plots are an indication that there would be kinks in twisting rigidity as contact regimes between scales undergo change as depicted, for instance in Fig.~\ref{fig:different modes of contact}, inset C to D. This plot can also be used to note the sensitivity of twisting to existing bending strains. We find that the sensitivity is quite high and thus any unintended bending either due to sample processing or loading asymmetries can drastically change the overall kinematics. Another remarkable feature is that, locking angle for the scale seems to be insensitive to bending. Hence, the locking envelopes computed from pure twisting would still hold in this case. Overall, bending strains have a differential effect on the twisting kinematics, affecting the kinematics trajectory substantially while leaving the locked state unchanged. 

In the same spirit, in  Fig.~\ref{fig:kappa31 vs theta different kappa13} (b), we highlight the effect of twisting on the bending behavior of the biomimetic scale beam. Here, we find that a twist in the system also has a significant impact on the bending behavior. With increasing positive twist, the bending engagement occurs earlier, whereas the opposite occurs with negative twist. This trend reflects the fact that twisting on one side has an `opening' effect on the scale, where they move apart from each other compared to the other side which leads to a 'closing' effect. The slopes also change significantly indicating a potential effect on the overall stiffness of the system. The abrupt changes correspond to sudden changes in contact regimes and shown with different colors. Here again, interestingly the presence of twist does not affect the locking angle, analogous to the twisting case.   


The validation of these analytical results using FE models are discussed next. Fig.~\ref{fig:kappa31 vs theta different kappa13} (c) and (d) are the counter parts of Fig.~\ref{fig:kappa31 vs theta different kappa13} (a) and (b), respectively. In Fig.~\ref{fig:kappa31 vs theta different kappa13} (c) we compare the FE results (shown as dots) with the results of our model (shown as dashed lines) for various values of pre-set bending strains ($\hat\kappa_1$) in the substrate. The kinks seen in the curves (close to $\hat\kappa_3=0.05$) are precisely the transition from red to green curves shown in Fig.~\ref{fig:kappa31 vs theta different kappa13} when the mode of contact changes. Analogous results for $\theta$ versus $\hat{\kappa}_1$ for different values of pre-set twisting strains $\hat\kappa_3$ are shown in Fig.~\ref{fig:kappa31 vs theta different kappa13} (d). Note that the pre-set strains for the two set of simulations approximate. This is because when the second quasi-static step (see Sec. \ref{sec:fem}) is performed, the strain values fixed in during the first quasi-static step does not remain fixed. That is, as we twist a pre-bent beam, the bending curvatures change mildly under the twisting deformation. In Fig.~\ref{fig:kappa31 vs theta different kappa13} (c) and (d), we account for this change by interpolating over the changing values of $\hat{\kappa}_1$ and $\hat{\kappa}_3$, respectively.

\begin{figure}[h!]
 \centering
 \begin{tabular}{cc}
 \includegraphics[width=2.9in]{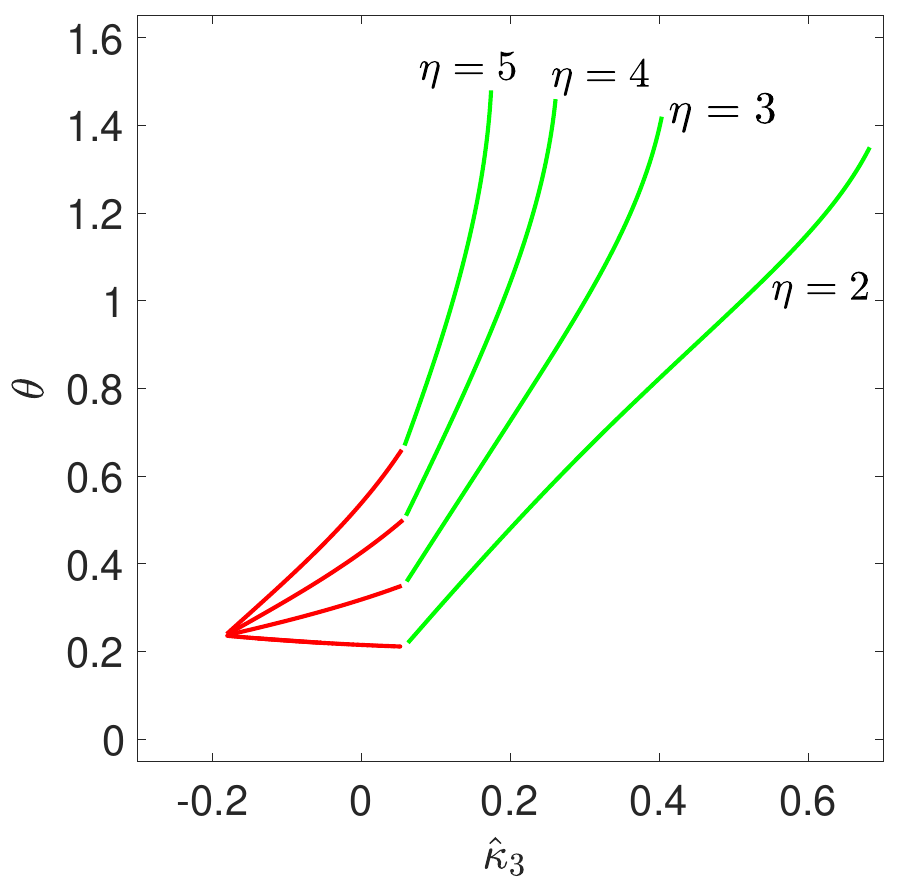}
 & \includegraphics[width=2.9in]{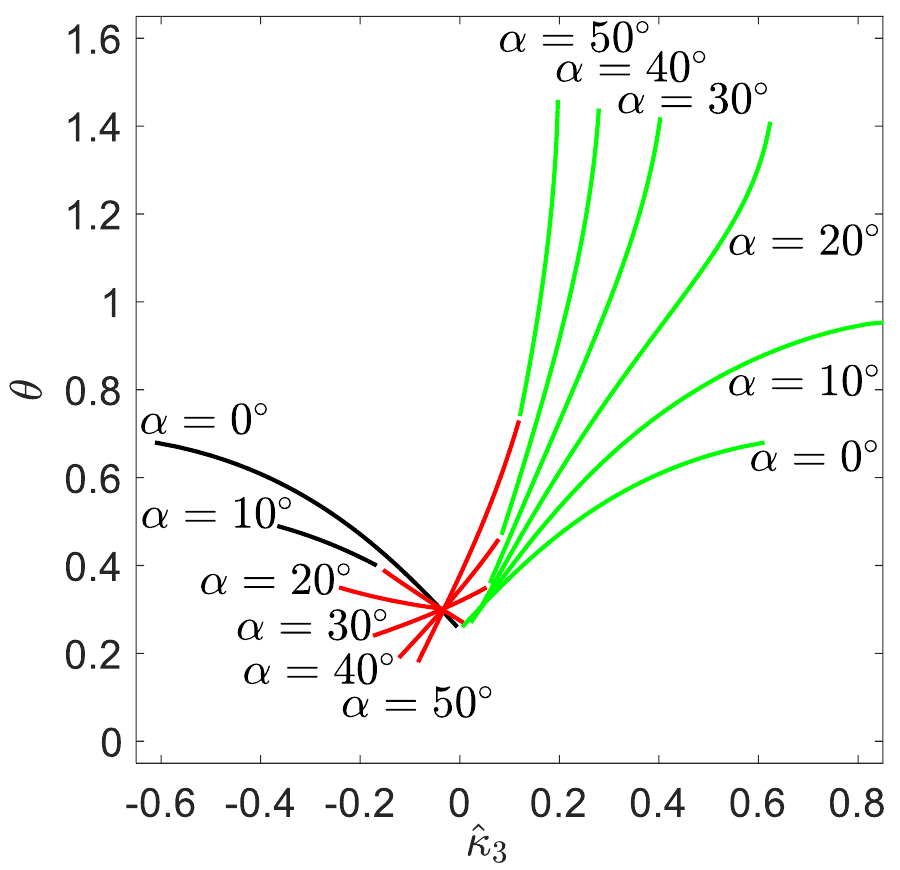}\\
 (a) & (b)\\
 \includegraphics[width=2.9in]{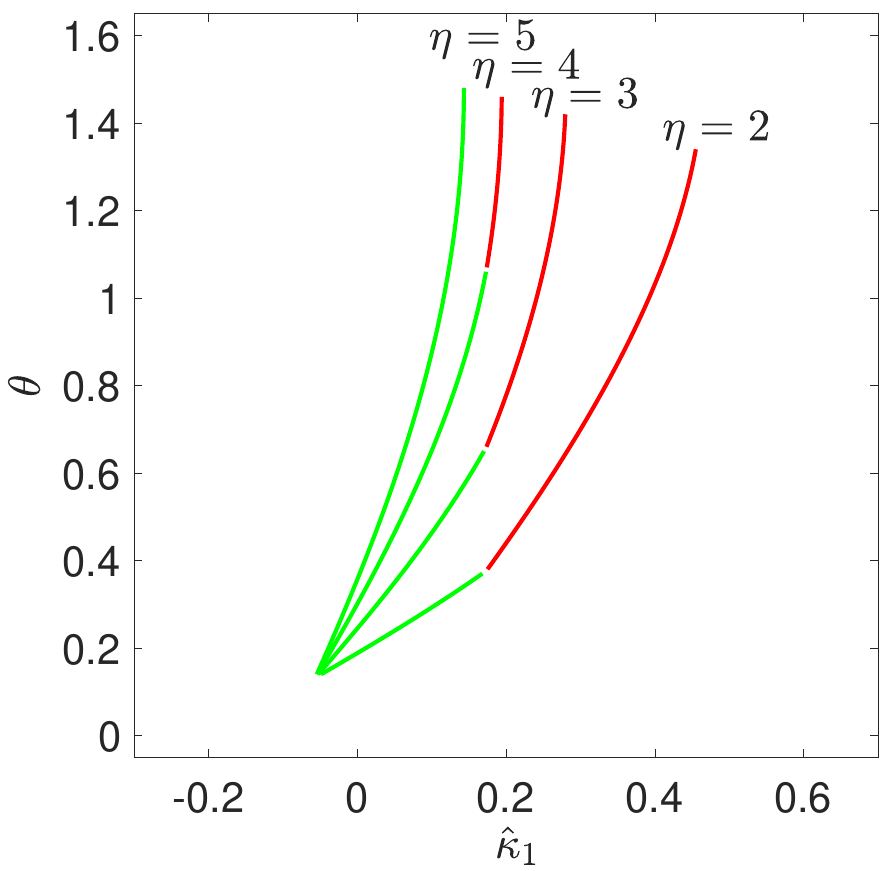}
 & \includegraphics[width=2.9in]{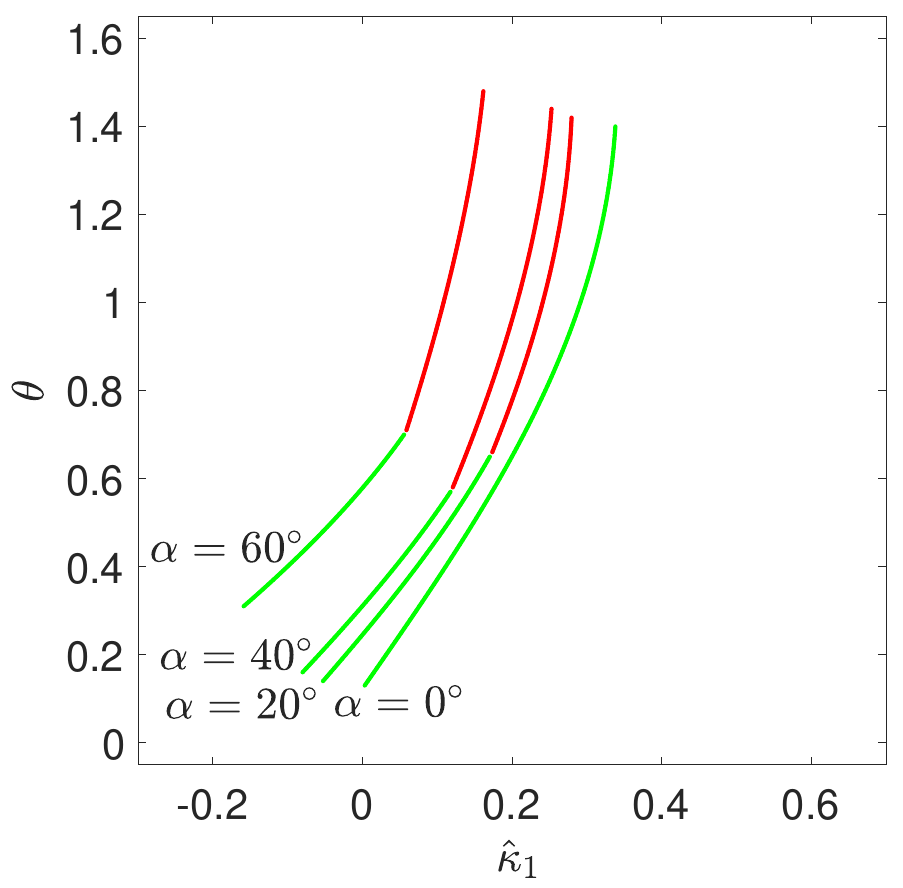}\\
 (c) & (d)
\end{tabular}
 \caption{(a) $\theta$ versus $\hat{\kappa}_3$ for different values of $\eta$ (for $\beta=1.25,\;\alpha=30^\circ,\;\hat{\kappa}_1=0.1$). (b) $\theta$ versus $\kappa_3$ for different values of $\alpha$ ($\eta=3,\;\beta=1.25$, $\kappa_1=0.1$). (c) $\theta$ versus $\kappa_1$ for different values of $\eta$ (for $\beta=1.25,\;\alpha=30^\circ,\;\kappa_3=0.1$). (d) $\theta$ versus $\kappa_1$ for different values of $\alpha$ ($\eta=3,\;\beta=1.25$, $\kappa_3=0.1$).}
 \label{fig:effect alpha and eta}
\end{figure}

\begin{figure}
    \centering
    \includegraphics[width=3.3in]{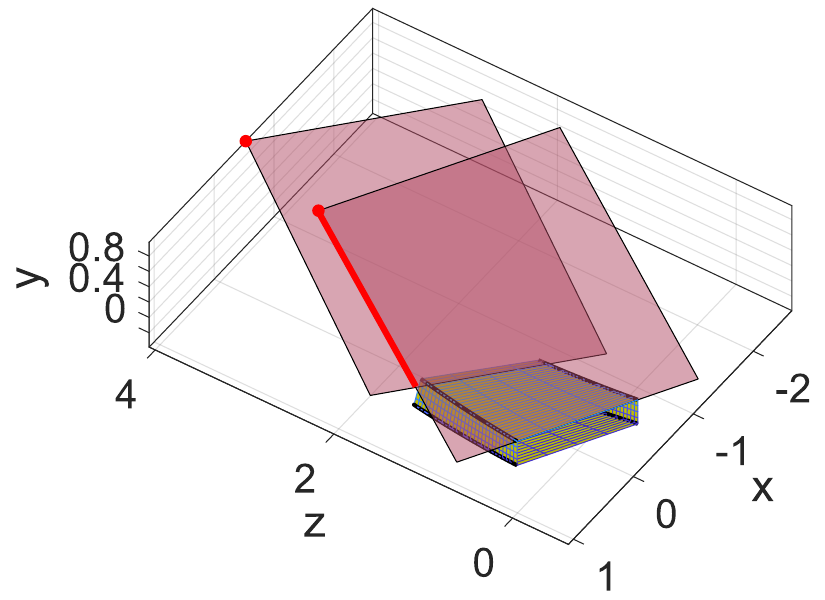}
    \caption{Visual depiction of the scales in contact at the point of disengagement as scales 'open' up at $\kappa_3=\kappa_3^*$. (for $\beta=1.25,\;\alpha=30^\circ,\;\hat{\kappa}_1=0.1$)}
    \label{fig:edge_face_convergence}
\end{figure}

We now use our model to explore the twisting kinematics at a given bending strain, with changing geometric parameters, $\alpha$ and $\eta$. These parameters have shown to be critical in dictating the overall kinematics of pure twisting~\cite{ebrahimi2019tailorable}. For this, we fix the bending curvature at $\hat{\kappa}_1=0.1$. 
First, we probe the significance of overlap ratio $\eta$ on the overall kinematics of the system. In Fig.~\ref{fig:effect alpha and eta} (a), we plot $\theta$ vs $\hat{\kappa}_3$ for different values of $\eta$ (for $\beta=1.25$, $\alpha=30^\circ$ and $\kappa_1=0.15$). 
Here, we first note the similarities with pure twisting case. Like pure twisting, increasing $\eta$ leads to steeper slopes in the $\theta-\hat{\kappa}_3$ plots. However, the twist completely changes the kinematic trajectories by introducing two distinct contact regimes. The abrupt changes in contact are more apparent for lower overlap ratio but are ameliorated at higher overlaps. Note that on the right side, the plots continue to increase in steepness until the locked states are reached. On the other hand, on the left side (negative twist), the decrease in slope (flatten) till they collapse into a point. This is the point of disengagement ($\kappa_3=\kappa_3^*$) of the scales beyond which, they no longer remain in contact. A surprising outcome of this plot is that, the disengagement point seems to be independent of the overlap ratio. This can be understood if this point is visualized for an RVE, Fig.~\ref{fig:edge_face_convergence} (intersection between the two scales is shown in red). Evidently, at this point, the nature of contact is the entire edge of one scale moving over the adjacent one.  Thus, at $\kappa_3^*$,  $\eta$ itself is not well defined and the scale angle is purely determined by the dihedral angle of two adjacent intersecting planes that is dependent only on the applied strains. 


We see here again that $\eta$ influences the angle of locking. We find that this parameter has a significant role to play in determining the overall kinematics of the system. Increasing $\eta$ leads to an overall stiffer nonlinearity but at the same time can potentially change the overall contact regime sequence up until locking.
Next, In Fig.~\ref{fig:effect alpha and eta} (b), we plot $\theta$ vs $\hat{\kappa}_3$ for different values of $\alpha$ (for $\beta=1.25$, $\eta=3$ and $\hat{\kappa}_1=0.1$). The effect of the tilt angle $\alpha$ is dramatic. For relatively small angles, i.e. $\alpha<20^\circ$, there is a relatively `stiff' response on either direction of twisting. These reflect scales sliding on either direction of twisting. However, there is a marked anisotropy between the directions in both magnitude and contact regimes. More interestingly, at higher tilt angles such bi-directionally distinct stiffness disappears altogether indicating the scales do not or only weakly engage in other other direction. The lack of engagement in the other direction can be visualized as `opening' of the scales in one direction vs closing. The effect of $\alpha$ on locking is also pronounced. 
At lower tilt angles, the stiffening effect disappears into a lockless behavior. This is due to scales sliding past each other without ever satisfying the contact constraints. Locking behavior emerges again for higher tilt angles. 

In Fig.~\ref{fig:effect alpha and eta} (c), we now look at the sensitivity of the bending kinematics in the presence of twist. For these plots the we keep the geometry of the substrates the same and now impose a twist of $\hat{\kappa}_3=0.1$ before bending and keep the tilt angle $\alpha=30^\circ$. Here, we see that the overall impact of higher $\eta$ is to increase the slope of $\theta-\hat{\kappa}_1$ plot, similar to pure bending~\cite{ghosh2014contact}. However, the presence of twist changes the nature of the $\theta-\hat{\kappa}_1$ curve by introducing different contact regimes, very similar to the twisting case discussed earlier (Fig.~\ref{fig:effect alpha and eta} (a)). Similar to that case, the abruptness of the kinks diminish at higher $\eta$, disappearing altogether when $\eta=5$. Thus, higher scale overlap has a suppressing effect on contact regime transitions. Interestingly, the scales remain engaged even at a negative bending value, due to the `closing' effect on scales due to twisting where scales are pushed closer to each other. However, eventually the scales again lose contact as bending on the other direction increases sufficiently, once again exhibiting the 'convergence' at the point of disengagement described previously.  

 We also investigate the effect of an existing twist on $\theta-\hat{\kappa}_1$ relationship for various tilt angles $\alpha$ but with fixed $\eta=3$ in Fig.~\ref{fig:effect alpha and eta} (d). Here, we find that at lower tilt angle, the contact regime transition is absent, appearing as tilt angle increases. This shows that the tilt angle is an important parameter that controls the complexity of contact regimes. Overall, the effect of pre-twist in bending is muted at lower tilt angles and higher overlap and more pronounced otherwise.

\begin{figure}[h!]
    \centering
    \includegraphics[width=3.3in]{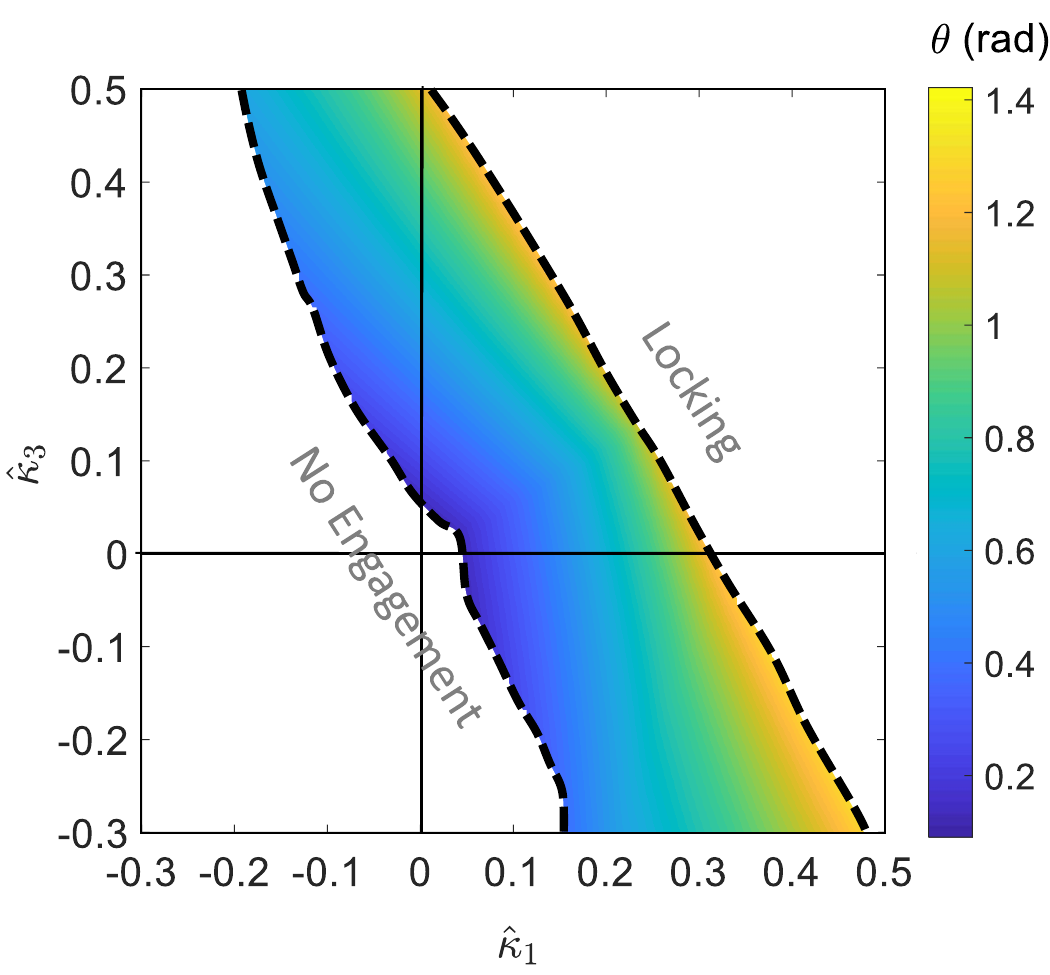}
    \caption{Slice of the engagement region, $\Gamma_e$, as defined in \eqref{eq:escale} for $\hat{\kappa}_2=0$. Points lying inside the shaded region correspond to configurations where scales are in contact.  The color bar shows the scale angle $\theta$ (in radians) for the corresponding configuration. Here $\theta_0=5^\circ$ and $\alpha=30^\circ$, $\eta=3$, and $\beta=1.25$.}
    \label{fig:engagement region}
\end{figure}

The investigation above, set the stage to investigate the positive-negative effects of twist-bend combinations. To put this bend-twist kinematics in more general context, we compute the curvature dependence of scale angles for a biomimetic scale beam of fixed geometric parameters. This results in a phase plot as shown in Fig.~\ref{fig:engagement region}. The curvature limits are taken to maintain self consistency with respect to contact regimes studied. For higher twist levels, more contact possibilities would need to be included. The phase plot in Fig.~\ref{fig:engagement region} is by solving the kinematic relationships developed earlier connecting scale angles and curvatures numerically at various values of bending and twisting strains. 

The phase boundary on the left is the limit of engagement, i.e. we do not have any scales engagement beyond this boundary. The phase boundary on the right hand side is the locking boundary. Any $(\hat{\kappa}_1,\hat{\kappa}_3)$ combination is forbidden in this region. Thus, the region in between indicates the region of engagement. It is interesting to note that there is no inherent symmetry between bending and twisting. In other words, the effect of bending on twisting is fundamentally different from the converse. In addition, one can see that multiple bend-twist combinations can give rise to same scale angles. This angular degeneracy is quite remarkable and could have significance for inverse designs. 

\begin{figure}[htbp]
    \centering
    \begin{tabular}{cc}
\includegraphics[width=3in]{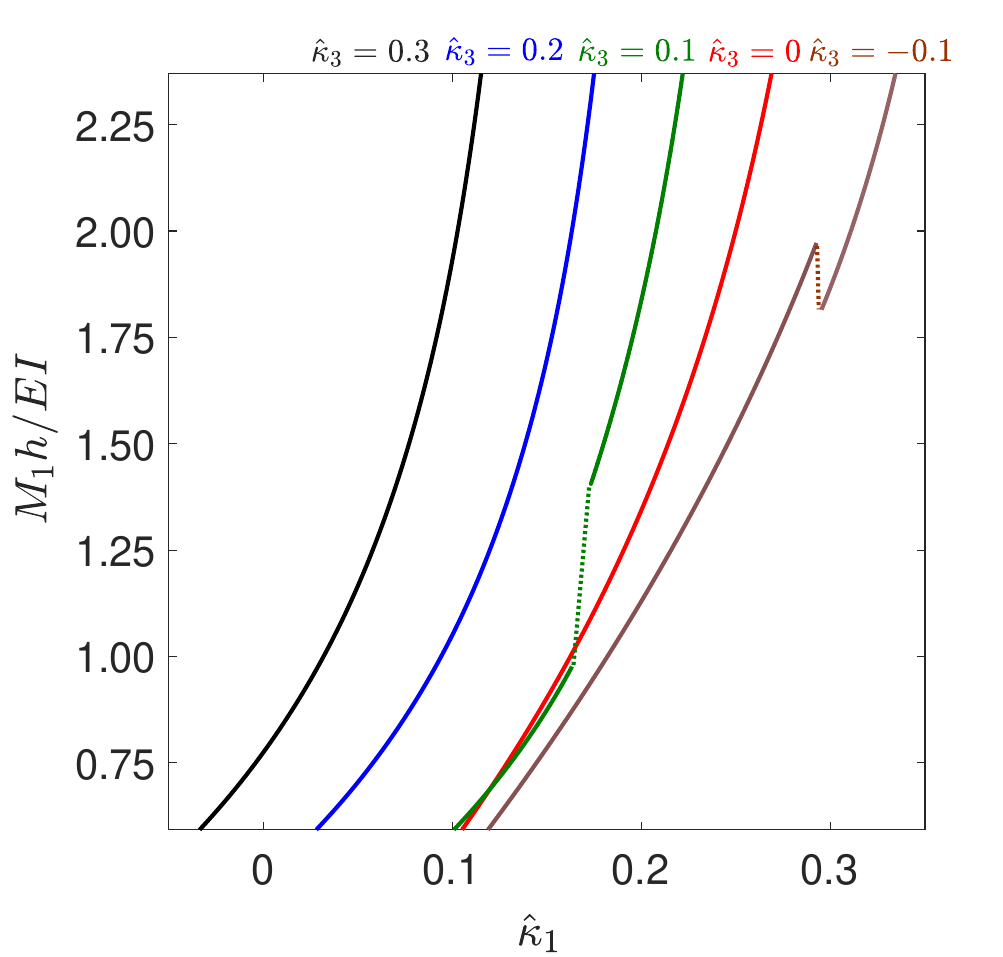} &
\includegraphics[width=2.9in]{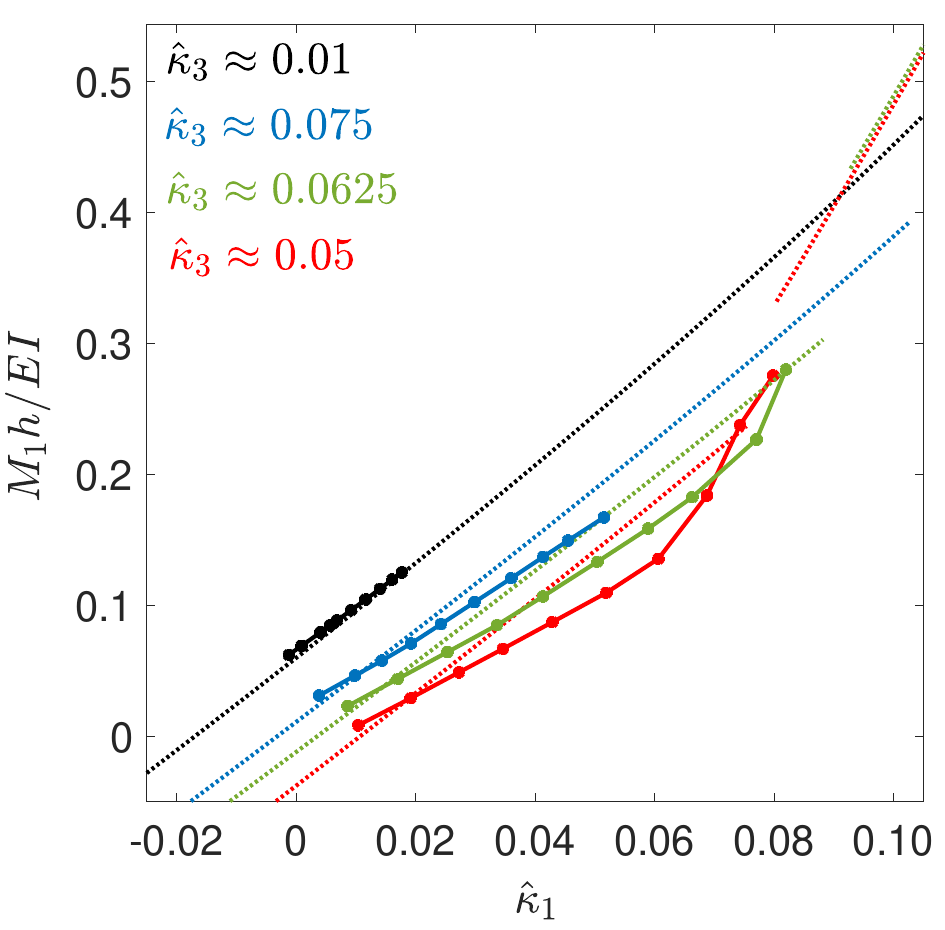} \\
  (a)   & (b)
\end{tabular}
    \caption{Bending moment $M_1$ versus $\hat{\kappa}_1$ for various values of $\hat{\kappa}_3$: (a) Plots for the analytical model (b) Plots using FE analysis with dashed lines showing theoretical predictions.}
    \label{fig:bending_moment}
\end{figure}

Finally, we discuss the mechanical behavior of these systems. In Fig.~\ref{fig:bending_moment} (a) we plot bending moment vs bending strains and discover the role played by the presence of twist for a given $\eta$. Higher values of positive twist shift the engagement to earlier parts, whereas higher values of negative twist shift it in the opposite direction.  
Interestingly, unlike smooth plots seen for pure bending earlier~\cite{ghosh2014contact}, twist effectively changes the contact regime in terms of discontinuities and jumps in the plot indicating sudden changes in bending rigidity. When a twist is added in the positive direction, the neutral position (no bending strain) is fully engaged (due to the engagement brought about by the twist). The FE comparison with these plots is shown in Fig.~\ref{fig:bending_moment} (b). Notice the validation region is for much smaller values of bending and twisting. This is due to inherent limitation of traditional FE software in simulating this type of system and already well known in literature ~\cite{ghosh2014contact,ebrahimi2019tailorable,shafiei2021very}. The agreements are excellent with respect to the theoretical values. The observed deviations are expected both due to the complex nature of contact, imposing global periodicity, and also the difficulties in keeping one curvature constant in an actual numerical simulation due to edge effects. 

\begin{figure}[h!]
    \centering
    \begin{tabular}{cc}
\includegraphics[width=2.9in]{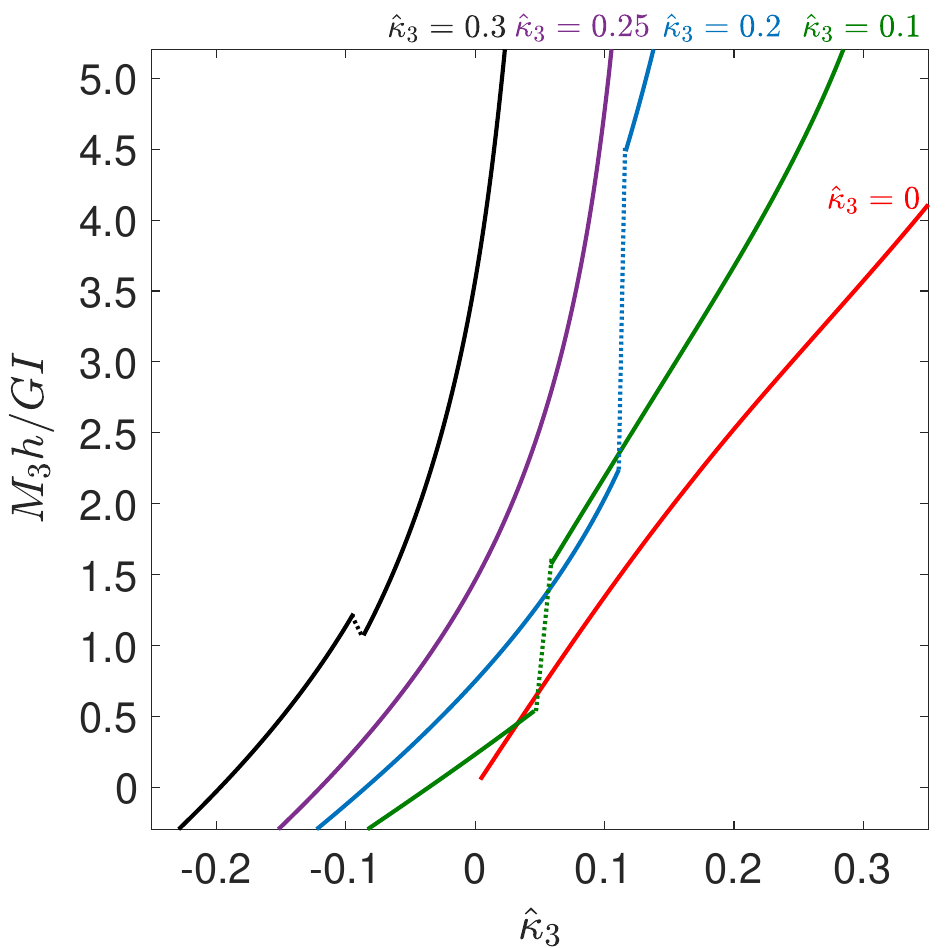} &
\includegraphics[width=2.9in]{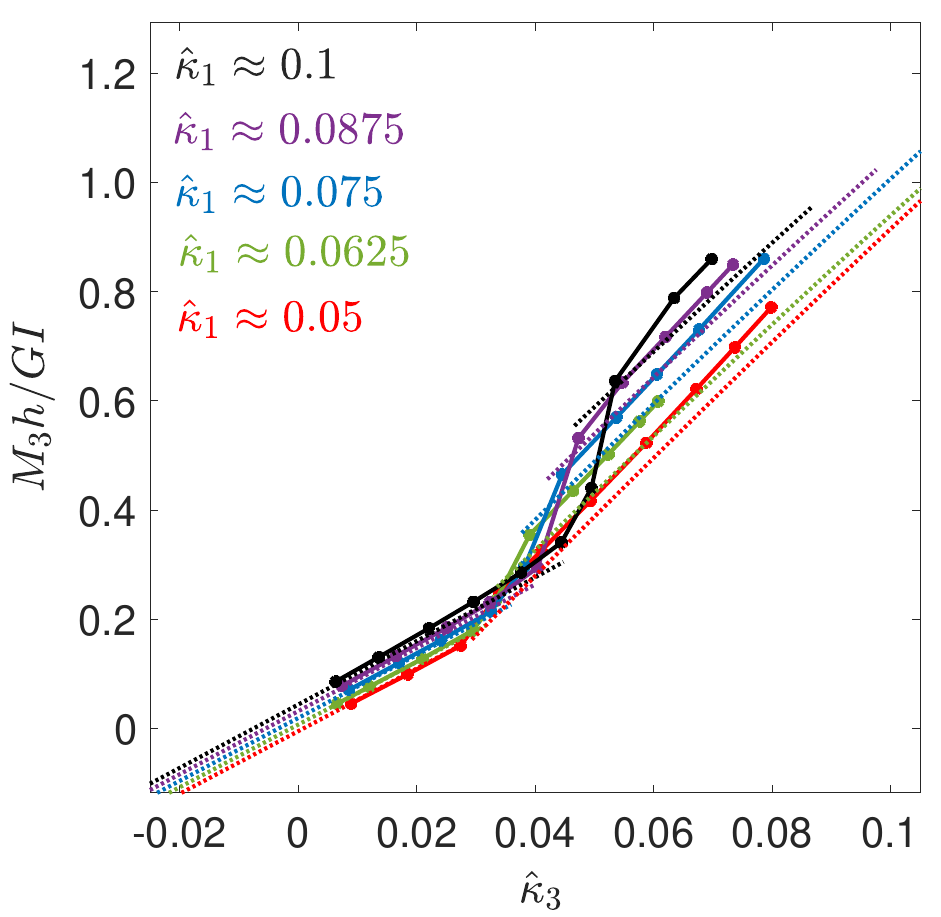} \\
  (a)   & (b)
\end{tabular}
    \caption{Twisting moment $M_3$ versus $\hat{\kappa}_3$ for various values of $\hat{\kappa}_1$: (a) Plots for the analytical model (b) Plots using FE analysis with dashed lines showing theoretical predictions.}
    \label{fig:torsion_moment}
\end{figure}

We see similar and even more dramatic effect in the twist-torque diagram in Fig.~\ref{fig:torsion_moment} (a). Here again, we see the  role of the bending strains as the torque-twists which are smooth nonlinear plots in pure twisting case~\cite{ebrahimi2019tailorable} now turn into highly discontinuous plots with disparate twist modulus. Yet again, an existing bending strain can cause the neutral position (zero twist strain) to be pre-engaged due to bending. These highlight the appreciable differences that can be brought about from cross-coupling effects.  Fig.~\ref{fig:torsion_moment} (b) shows the companion FE torque-twist plots that also indicate excellent agreement with the analytical results. As in the case for bending, the deviations are mainly because of the difficulties in keeping the cross curvatures fixed for the duration of the simulation. The other sources of errors are possibly due to assuming fixed curvatures as opposed to spatially varying curvatures arising in a FE simulation. Analogous to bending case, imposing global periodicity in scale engagements also leads to some differences in behavior. Please note that many of these non ideal effects and sources of discrepancies have been discussed earlier ~\cite{ghosh2017non,ali2019bending}.

\section{Conclusions}
In this work, we addressed the cross coupling effects of bending and twisting in a biomimetic scale elastic beam for the first time. Here the scales were plate like rectangular inclusions protruding at an angle from the surface of the elastic substrate. We find highly intricate and often surprising effect of one over the other across the kinematics and mechanics. We quantified these effects by developing analytical relationships within the framework of Cosserat kinematics and global-local energy balance. This model reduced to the earlier developed model for pure bending and twisting in literature and was also validated with FE simulations. This study completes a significant missing piece in the mechanics of fish scale inspired biomimetic system, which is of great practical importance in applications ranging from high performance smart skins to soft robotic systems. 

\section*{Acknowledgement}
This work was supported by the United States National Science Foundation’s Civil, Mechanical, and Manufacturing Innovation, CAREER Award \#1943886.

\bibliographystyle{elsarticle-num} 
\bibliography{bibliography}

\appendix 

\section{Extracting Cosserat Strains from FE Data}
\label{AppendixA}

To compare the FE simulations with the analytical model presented in Sec. \ref{sec:engagement}, we extract the Cosserat bending strains $\kappa_1$, $\kappa_2$, and $\kappa_3$ from the FE data through the following steps. First, we extract the position vector for the mid-point of the top face of the beam from the numerical simulations. This vector is an estimate for $\mathbf{r}(s)$ as given in \eqref{eq:r=int}. The information of the directors $\mathbf{d}_1(s)$ and $\mathbf{d}_2(s)$ is extracted by subtracting the position vectors of center-line and the right-edge of the beam, respectively, from the estimate for $\mathbf{r}(s)$ and normalizing the resulting quantities to produce unit vectors. After validating inextensibility by verifying that the change in length of the rod is small (with typical percentage relative error $\Delta L_B/L_B \approx 1\%$), we compute the third director using  $\mathbf{d}_3=\mathbf{r}'(s)$, estimating the right hand side using finite differences along the beam. It follows from \eqref{eq:d=Qe} that rotation matrix mapping the cross-sections of the beam is given by $\mathbf{Q}(s)=[\mathbf{d}_1(s), \mathbf{d}_2(s), \mathbf{d}_3(s)]$ where the directors are taken to be the column vectors of the matrix. The skew-symmetric matrix $\mathbf{K}$ containing the bending strains $\kappa_1(s),\;\kappa_2(s),\;\kappa_3(s)$ (cf. \eqref{eq:Kkappa}) along the length of the beam is computed using the formula  $\mathbf{K}=\mathbf{Q}'(s)\mathbf{Q}^{-1}(s)$, where $\mathbf{Q}'(s)$ is estimated using finite differences. To avoid boundary effects, we average the bending strains over the middle half of the beam to obtain estimates for average bending strains in the beam. 

\begin{figure}[htbp]
    \centering
    \includegraphics[width=4in]{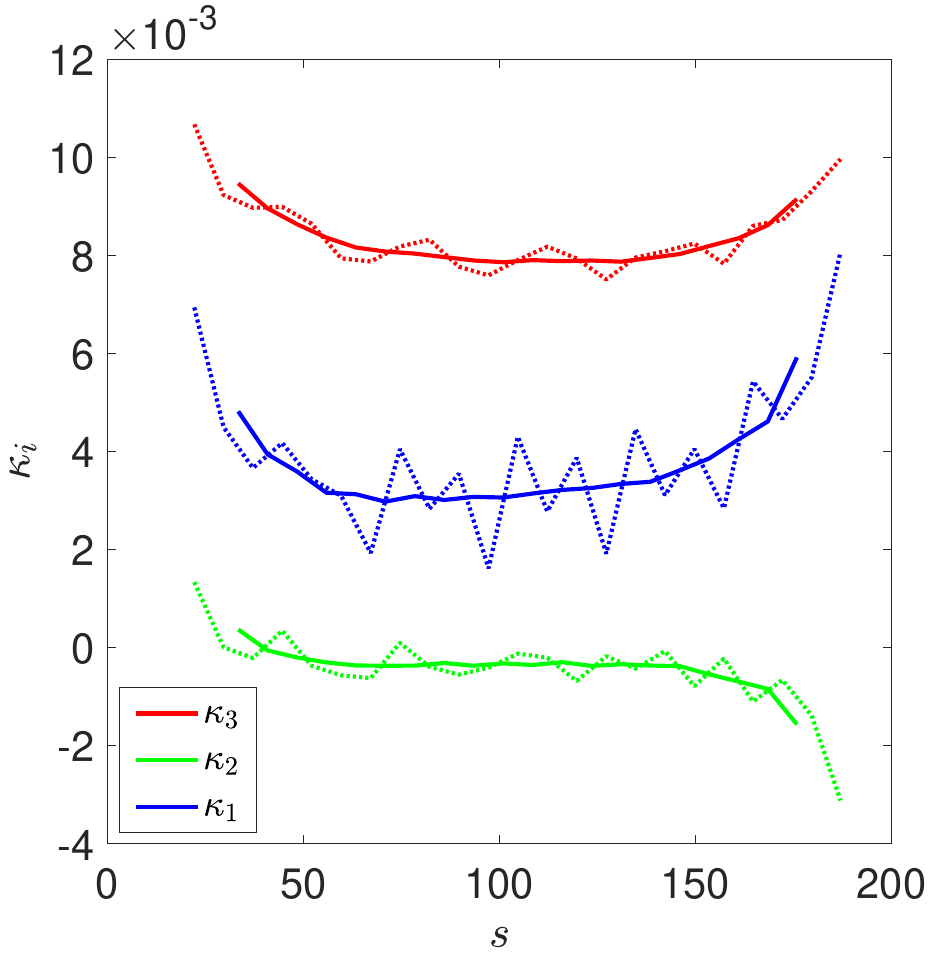}
    \caption{Extracting Cosserat strain from FE Data.}
    \label{fig:extracting strain}
\end{figure}
\end{document}